\documentclass[12pt,a4paper]{article}
\usepackage{graphicx}
\usepackage[T1]{fontenc}
\usepackage[utf8]{inputenc}
\usepackage{textcomp}
\usepackage[sc,osf]{mathpazo}
\usepackage{a4wide}  
\usepackage{latexsym,amsthm,amsfonts,amsmath,mathrsfs,amssymb}
\usepackage{dsfont}
\usepackage{accents}
\usepackage[nosort]{cite}
\usepackage{booktabs} 
\usepackage[unicode,implicit]{hyperref}
\hypersetup{%
  pdftitle    = {Wald entropy in Kaluza-Klein black holes}
  pdfkeywords = {Black holes, thermodynamics, Noether charge, Wald entropy,
    first law, magnetic charge, dimensional reduction, Kaluza-Klein theory, compactification},
  pdfauthor   = {Carmen G\'omez-Fayr\'en, Patrick Meessen, Tom\'as Ort\'{\i}n, Matteo Zatti},
  plainpages  = true,
  colorlinks  = true,
  citecolor   = blue,
  urlcolor    = red,
  linkcolor   = black
}
\newcommand{\hepth}[1]{{\tt
\href{http://www.arXiv.org/abs/hep-th/#1}{hep-th/#1}}}
\newcommand{\grqc}[1]{{\tt
\href{http://www.arXiv.org/abs/gr-qc/#1}{gr-qc/#1}}}

\newcommand{\arxiv}[1]{{\tt arXiv:\href{http://www.arXiv.org/abs/#1}{#1}}}
\newcommand{\doi}[1]{{\tt DOI:\href{http://dx.doi.org/#1}{#1}}}

\allowdisplaybreaks

\makeatletter
\@addtoreset{equation}{section}
\makeatother

\begin{document}

\begin{flushright}
\small
IFT-UAM/CSIC-23-10\\
May 2\textsuperscript{nd}, 2023\\
\normalsize
\end{flushright}

\vspace{1cm}

\begin{center}

  {\Large {\bf Wald entropy in Kaluza-Klein black holes}}

\vspace{1.5cm}

\renewcommand{\thefootnote}{\alph{footnote}}

{\sl\large Carmen G\'omez-Fayr\'en},$^{1,}$\footnote{Email: {\tt carmen.gomez-fayren[at]estudiante.uam.es}}
{\sl\large Patrick Meessen,$^{2,3,}$}\footnote{Email: {\tt meessenpatrick[at]uniovi.es}}
{\sl\large Tom\'as Ort\'{\i}n},$^{1,}$\footnote{Email: {\tt  tomas.ortin[at]csic.es}}\\[.5cm]
{\sl\large and Matteo Zatti$^{1,}$}\footnote{Email: {\tt matteo.zatti[at]estudiante.uam.es}}

\setcounter{footnote}{0}
\renewcommand{\thefootnote}{\arabic{footnote}}
\vspace{1cm}

${}^{1}${\it Instituto de F\'{\i}sica Te\'orica UAM/CSIC\\
C/ Nicol\'as Cabrera, 13--15,  C.U.~Cantoblanco, E-28049 Madrid, Spain}\\

\vspace{0.2cm}

${}^{2}${\it HEP Theory Group, Departamento de F\'{\i}sica, Universidad de Oviedo\\
  Avda.~Calvo Sotelo s/n, E-33007 Oviedo, Spain}\\

\vspace{0.2cm}

${}^{3}${\it Instituto Universitario de Ciencias y Tecnolog\'{\i}as Espaciales
  de Asturias (ICTEA)\\ Calle de la Independencia, 13, E-33004 Oviedo, Spain}

\vspace{1cm}


{\bf Abstract}
\end{center}
\begin{quotation}
  {\small We study the thermodynamics of the 4-dimensional electrically
    charged black-hole solutions of the simplest 5-dimensional Kaluza-Klein
    theory using Wald's formalism. We show how the electric work term present
    in the 4-dimensional first law of black-hole thermodynamics arises in the
    purely gravitational 5-dimensional framework. In particular, we find an
    interesting geometric interpretation of the 4-dimensional electrostatic
    potential similar to the angular velocity in rotating black
    holes. Furthermore, we show how the momentum map equation arises from
    demanding compatibility between the timelike Killing vector of the
    black-hole solution and the spatial Killing vector of the 5-dimensional
    background.}
\end{quotation}

\newpage
\pagestyle{plain}

\tableofcontents


\section*{Introduction}

The possible existence of additional dimensions of different kinds has been a
recurring theme in theoretical physics for most of the past century since the
first proposal by Kaluza and Klein \cite{kn:Kal,kn:Kle}.\footnote{These and
  many other articles on Kaluza-Klein theories can be found in
  Ref.~\cite{Appelquist:1987nr}.} In Kaluza-Klein (KK) theories, gravity with
additional spatial dimensions gives rise to gauge and scalar fields, providing
a framework for the unification of gravity with the rest of the fundamental
interactions, which are described by Abelian and non-Abelian Yang-Mills- (YM)
type fields. Actually, according to Ref.~\cite{Straumann:2000zc} it seems that
the latter were discovered by Pauli precisely in this framework in which they
arise quite naturally because, after all, YM theories are nothing but
geometric theories with extra dimensions corresponding to the fibers of the
corresponding principal bundle.\footnote{See, for instance, the classical
  review Ref.~\cite{Eguchi:1980jx}.}

This aspect of YM theories as theories of extra dimensions is often neglected
because, when they are not the result of a KK theory, those extra dimensions
are not standard spacetime dimensions and cannot be ``seen'' by the metric
field that describes gravity as it happens in KK theories. It is this
separation between spacetime dimensions that can be seen by the metric and
extra, ``gauge'', dimensions, that cannot, that leads to the mathematical
structure of principal bundle to describe the geometry of YM theories. In the
context of KK theories, all dimensions can be treated in the same fashion and
the simpler structure of a simple (pseudo-) Riemannian differential manifold
suffices to describe geometries which include YM fields but can also be
simultaneously described in terms of principal bundles.

Using this dual description of KK geometries as (pseudo-) Riemannian manifolds
or as principal bundles one should be able to prove results using the former
formulation, which is simpler and more widely understood. One of the main
goals of this paper is to use this mechanism to study the spacetime symmetries
of systems with YM fields in terms of the symmetries of (pseudo-) Riemannian
manifolds which, as is well known, are isometries generated by Killing
vectors. We will consider the simplest setting, 5-dimensional general
relativity in a manifolds with a single compact dimension which gives rise to
a U$(1)$ gauge field, but our results should apply to more general settings.

The study of the spacetime symmetries of gauge field configurations arises
naturally in the construction of spacetime conserved quantities
\cite{Barnich:2001jy,Barnich:2003xg} and, via Wald's discovery of the
connection between the Noether charge associated to diffeomorphisms and the
black hole entropy \cite{Lee:1990nz,Wald:1993nt}, in the context of black-hole
thermodynamics in presence of matter fields with gauge freedoms
\cite{Jacobson:2015uqa,Prabhu:2015vua,Elgood:2020svt,Elgood:2020mdx,Elgood:2020nls}. The
main observation is that one has to search for symmetries in the complete
bundle and that those symmetries, when seen from (or projected to) the base
space, are combinations of a diffeomorphism and a ``induced'' or
``compensating'' gauge transformation. This gauge transformation depends on the
diffeomorphism and cannot be ignored or separated from it. As a consequence,
most fields cannot be treated as simple tensors under diffeomorphisms as in
\cite{Iyer:1994ys}. This is a fundamental fact that we are going to prove
using the KK framework,\footnote{A purely principal-bundle-based approach can
  be found in \cite{Prabhu:2015vua}.} but one can arrive at this conclusion by
considering spinors in curved spacetime.

Spinors (and Lorentz tensors) are defined in appropriate bundles connected to
the tangent space on which local Lorentz transformations act. Usually, they
are treated as scalars under diffeomorphisms but it is not difficult to see
that this description is incorrect: let us consider spinor fields in Minkowski
spacetime and let us consider the effect of an infinitesimal global Lorentz
spacetime (\textit{i.e.}~not tangent space) transformation on the spinors with
parameter $\sigma^{ab}$. If they are treated as scalars they will transform as
such, that is\footnote{In this simple example we are working in Cartesian
  coordinates and we are not distinguishing between spacetime and tangent
  space indices.}

\begin{equation}
  \delta_{\sigma}\psi
  =
  -\pounds_{k_{\sigma}}\psi
  = -\imath_{k_{\sigma}}d\psi\,,
  \hspace{1cm}
  k_{\sigma}
  \equiv
  \sigma^{\mu}{}_{\nu}x^{\nu}\partial_{\mu}\,.
\end{equation}

\noindent
Thus, they will not transform as spinors under that spacetime transformation
as they certainly should under a Lorentz transformation.

The solution to th eabove problem comes from the following observation: the
spacetime diffeomorphism generated by the Killing vector $k_{(\sigma)}$ induces
a tangent space Lorentz transformation with a parameter that is minus the
(automatically antisymmetric ) derivative of the Killing vector, also known as
\textit{Killing bivector} or \textit{Lorentz momentum map}. In this case

\begin{equation}
-\partial_{\mu}k_{(\sigma) \nu}= \sigma_{\mu\nu}\,,
\end{equation}

\noindent
as expected. In more general settings the parameter of the induced local
Lorentz transformation includes a term proportional to the spin connection
\cite{Ortin:2002qb} and is, indeed, local.

The combination of the Lie derivative and the compensating Lorentz
transformation for the infinitesimal diffeomorphism generated by an arbitrary
Killing vector field $k$ is known as the \textit{spinorial Lie derivative}
$\mathbb{L}_{k}$ and was first introduced by Lichnerowicz and Kosmann in
Refs.~\cite{kn:Lich,kn:Kos,kn:Kos2} and later studied and extended in
Refs.~\cite{Hurley:cf,Vandyck:1988ei,Vandyck:1988gc,Ortin:2002qb} also as
the \textit{Lorentz-covariant Lie derivative} or as the \textit{Lie-Lorentz
  derivative}. One of its main properties is that it transforms covariantly
under further diffeomorphisms and local Lorentz transformations. Thus, the
invariance of the spinor field $\psi$ under the infinitesimal diffeomorphism
generated by $k$ reads

\begin{equation}
  \mathbb{L}_{k}\psi
  =
  0\,,  
\end{equation}

\noindent
and is invariant statement.\footnote{This and similar equations can be seen as
  equations determining the values of the vector fields and gauge parameters
  that, combined, leave invariant the fields, known as \textit{reducibility
    (or Killing) parameters} \cite{Barnich:2001jy}. Our approach stresses the
  invariance of the equations which is an important ingredient in the gauge
  invariance of the final results.}

Similar considerations lead to the definition of more general Lie covariant
derivatives \cite{Ortin:2015hya,Elgood:2020svt,Elgood:2020mdx,Elgood:2020nls}
with analogous properties. As stated above, one of our main goals is to test
the simplest of these constructions (the Lie-Maxwell derivative for U$(1)$
gauge fields) using the KK framework.

Another of our main goals in this paper is to study the thermodynamics of
4-dimensional KK black holes directly from the 5-dimensional point of view
using Wald's formalism. It is clear that we can only do this if we know the
relation between the 4- and 5-dimensional spacetime symmetries and this is one
of our motivations for its study. Furthermore, we know that the 4-dimensional
event horizon of stationary solutions is the Killing horizon of a certain
Killing vector but, does the existence of a 4-dimensional event horizon imply
the existence of a 5-dimensional one? Will it also be a Killing horizon? With
respect to which Killing vector? Will the 5-dimensional surface gravity be
equal to the 4-dimensional one? To the best of our knowledge, there are no
complete answers to these questions in the literature and we will try to find
them for static 4-dimensional black holes. 

Finally, we will apply all these results to the calculation of the Smarr
formula and first law in 5 dimensions, aiming to obtain those of the
4-dimensional black holes in a sort of ``dimensional reduction'' of those
formulae. We will succeed with the first but not with the second, which needs
further research.

This work is organized as follows: in Section~\ref{sec-KKtheory} we review the
most basic Kaluza-Klein theory, its dimensional reduction and symmetries. In
Section~\ref{sec-electricKKBH} we review the general electrically charged,
static solution of the 4-dimensional theory obtained by dimensional reduction
in the previous section.  In Section~\ref{sec-4dthermodynamics} we study the
thermodynamics of the black holes of the 4-dimensional theory using Wald's
formalism and the extensions necessary to account for the gauge symmetries. In
Section~\ref{sec-KKBHs5} we study 4-dimensional black holes from a
5-dimensional perspective and use the results to try rederive Smarr formula
and the first law Section~\ref{sec-5dthermodynamics}. Finally, in
Section~\ref{sec-discussion} we discuss our results and future directions of
work.

\section{Basic Kaluza-Klein theory}
\label{sec-KKtheory}

Consider pure Einstein gravity in 5 dimensions parametrized by the coordinates
$\hat{x}^{\hat{\mu}}$.\footnote{We write hats over all 5-dimensional objects
  to distinguish them from the 4-dimensional ones. The 5\textsuperscript{th}
  coordinate will be denoted as $x^{4} = z$ and the corresponding
  (\textit{world}) index will be $\underline{z}$ to distinguish it from the
  corresponding, not underlined, 5\textsuperscript{th} tangent space
  direction. Thus, $(\hat{\mu}) = (\mu,\underline{z})$, $(\hat{a})=(a,z)$,
  etc. We use a mostly minus signature and the rest of the conventions are
  those used in Ref.~\cite{Ortin:2015hya}.} The only dynamical field is the
5-dimensional metric $\hat{g}_{\hat{\mu}\hat{\nu}}$ and the 5-dimensional line
element is

\begin{equation}
  ds_{(5)}^{2}
  =
  \hat{g}_{\hat{\mu}\hat{\nu}}dx^{\hat{\mu}}x^{\hat{\nu}}\,.
\end{equation}

In this theory, the dynamics of the metric field is dictated by the
Einstein-Hilbert action

\begin{equation}
  \label{eq:5dEHaction}
  S[\hat{g}]
  =
  \frac{1}{16\pi G_{N}^{(5)}}\int d^{5}x\,\sqrt{|\hat{g}|}\, \hat{R}\,,
\end{equation}

\noindent
where $G_{N}^{(5)}$ is the 5-dimensional analog of the Newton constant.

If the 5\textsuperscript{th} coordinate is periodic

\begin{equation}
z\sim z+2\pi \ell\,,  
\end{equation}

\noindent
where $\ell$ is some length scale, all the components in the metric can be
expanded in Fourier series.\footnote{A common choice in the literature is
  $\ell=R$, the asymptotic radius of the compact dimension which will be
  defined below in Eq.~(\ref{eq:KKradius}). However, since that radius is
  invariant, as we are going to show, that choice (as well as the choice in
  which $z$ is an angle and $z\sim z+2\pi$) prevents any redefinition of the
  coordinate $z$. We are, however, specially interested in studying the
  rescalings of $z$ which induce a global rescaling symmetry of the
  4-dimensional theory and, therefore, we have to distinguish very carefully
  between $\ell$ and $R$. } Since the higher modes correspond to fields which
appear as massive from the non-compact 4-dimensional world perspective and
since their masses can be made arbitrarily high by choosing the size of the
5\textsuperscript{th} direction small enough, we can safely ignore them at low
energies and work with the zero modes only, which are the components of a
$z$-independent metric

\begin{equation}
  \label{eq:adaptedcoordinates}
\partial_{\underline{z}} \hat{g}_{\hat{\mu}\hat{\nu}}=0\,.  
\end{equation}

Thus, in this scenario the metric admits an isometry generated by a spacelike
Killing vector

\begin{equation}
  \hat{k}
  =
  \hat{k}^{\hat{\mu}}\partial_{\hat{\mu}}
  =
  \partial_{\underline{z}}\,,
\end{equation}

\noindent
and the coordinates we are using $(x^{\hat{\mu}})=(x^{\mu},x^{4}\equiv z)$ are
coordinates adapted to the isometry.

It is important to distinguish between the length scale $\ell$ and the radius
of the compact direction, which is a local quantity $R(x)$ since the geometry
we are describing is that of a S$^{1}$ fibered over a 4-dimensional base
space. In these adapted coordinates, we can define it in a
coordinate-independent way as the integral of the 1-form

\begin{equation}
\hat{n} \equiv
-\frac{\hat{k}_{\hat{\mu}}}{\sqrt{-\hat{k}^{2}}}d\hat{x}^{\hat{\mu}}\,,
\hspace{1cm}
\hat{n}^{2}=-1\,,
\end{equation}

\noindent
along the S$^{1}$ at any point of the 4-dimensional base space:

\begin{equation}
  \label{eq:KKradius}
  2\pi R(x)
  \equiv
  \int_{S^{1}} \hat{n}
  =
  \int_{0}^{2\pi\ell}dz
  \sqrt{-\hat{g}_{\underline{z}\underline{z}}}\,.
\end{equation}

The constant asymptotic value of $R(x)$ will be denoted by $R$ and  will be
assumed to be finite.

The 5-dimensional metric can be decomposed in terms of fields which transform
as 4-dimensional fields are expected to. There is a scalar (the Kaluza-Klein
(KK) scalar field), $k$, defined by 

\begin{equation}
  k^{2}
  \equiv
  -\hat{k}^{2} = -\hat{g}_{\underline{z}\underline{z}}\,,
\end{equation}

\noindent
and constrained to take strictly positive values; the KK vector field

\begin{equation}
  A_{\mu} \equiv
  \hat{g}_{\mu\underline{z}}/\hat{g}_{\underline{z}\underline{z}}\,,
  \hspace{1cm}
  A
  \equiv 
  A_{\mu}dx^{\mu}\,,
\end{equation}

\noindent
and the KK metric

\begin{equation}
  g_{\mu\nu}  
  \equiv
  \hat{g}_{\mu\nu}
  -\hat{g}_{\mu\underline{z}}\hat{g}_{\nu\underline{z}}/\hat{g}_{\underline{z}\underline{z}}\,,
  \hspace{1cm}
  ds_{(4)}^{2}
  =
  g_{\mu\nu}dx^{\mu}dx^{\nu}\,.
\end{equation}

The 5-dimensional line element can be rewritten in terms of the 4-dimensional
KK fields we have just defined as

\begin{equation}
  ds_{(5)}^{2}
   =
    ds_{(4)}^{2} -k^{2}\left(dz+A\right)^{2}\,.
\end{equation}

The particular definitions of the 4-dimensional fields $g_{\mu\nu},A_{\mu},k$
are justified by their behaviour under 5-dimensional reparametrizations that
respect the gauge choice in Eq.~(\ref{eq:adaptedcoordinates}) (coordinates
adapted to the isometry).

Most of these reparametrizations are generated by $z$-independent
5-dimensional vectors $\hat{\xi}^{\hat{\mu}}$ which act on the 5-dimensional
metric according to

\begin{equation}
  \delta_{\hat{\xi}}\hat{g}_{\hat{\mu}\hat{\nu}}
  =
  -\pounds_{\hat{\xi}}\hat{g}_{\hat{\mu}\hat{\nu}}
  =
  -\left(\hat{\xi}^{\hat{\rho}}\partial_{\hat{\rho}}\hat{g}_{\hat{\mu}\hat{\nu}}
    +2\partial_{(\hat{\mu}}\hat{\xi}^{\hat{\rho}}\hat{g}_{\hat{\nu})\hat{\rho}}
  \right)\,.
\end{equation}

\noindent
It follows that their action on the 4-dimensional fields is 

\begin{subequations}
  \begin{align}
    \delta_{\hat{\xi}} k
    & =
      -\hat{\xi}^{\rho}\partial_{\rho}k\,,
    \\
    & \nonumber \\
    \delta_{\hat{\xi}}A_{\mu}
    & =
      -\left(\hat{\xi}^{\rho}\partial_{\rho} A_{\mu}
      +\partial_{\mu}\hat{\xi}^{\rho} A_{\rho}\right)
      -\partial_{\mu}\hat{\xi}^{\underline{z}}\,,
    \\
    & \nonumber \\
    \delta_{\hat{\xi}}g_{\mu\nu}
    & =
      -\left(\hat{\xi}^{\rho}\partial_{\rho}g_{\mu\nu}
      +2\partial_{(\mu}\hat{\xi}^{\rho}g_{\nu)\rho}
      \right)\,.
  \end{align}
\end{subequations}

\noindent
These transformations can be interpreted as 4-dimensional general coordinate
transformations generated by the 4-dimensional vector

\begin{equation}
\xi^{\mu}\equiv \hat{\xi}^{\mu}\,,
\end{equation}

\noindent
plus standard gauge transformations generated by the gauge parameter

\begin{equation}
\chi \equiv -\hat{\xi}^{\underline{z}}\,,
\end{equation}

\noindent
acting on $A$ only. Therefore, $A$ plays the role of a 1-form connection with
gauge transformations

\begin{equation}
\delta_{\chi}A = d\chi\,,  
\end{equation}

\noindent
and gauge-invariant field strength

\begin{equation}
  F= dA = \tfrac{1}{2}F_{\mu\nu}dx^{\mu}\wedge dx^{\nu}\,,
  \hspace{1cm}
  F_{\mu\nu} = 2\partial_{[\mu}A_{\nu]}\,.  
\end{equation}

There is only one $z$-dependent 5-dimensional general coordinate
transformation that preserves the gauge Eq.~(\ref{eq:adaptedcoordinates}). It
is generated by the vector field

\begin{equation}
  \hat{\eta}
  \equiv
  z\partial_{\underline{z}}\,,
\end{equation}

\noindent
and it only acts on the $z$ coordinate as a rescaling: if $\alpha$ is an
infinitesimal parameter, then

\begin{equation}
  \delta_{\alpha}z = \alpha z\,,
  \hspace{1cm}
  z' = e^{\alpha} z\,.  
\end{equation}

\noindent
Thus, if $z\in [0,2\pi \ell]$, $z' \in [0,2\pi \ell']$ with
$\ell' = e^{\alpha}\ell$. Let us stress that this transformation does not
change $R$.

This transformation only acts on components of the metric with a $z$ index:

\begin{equation}
  \delta_{\hat{\eta}} \hat{g}_{\mu\underline{z}}
  =
  -\hat{g}_{\mu\underline{z}}\,,
  \hspace{1cm}
  \delta_{\hat{\eta}} \hat{g}_{\underline{z}\underline{z}}
  =
  -2\hat{g}_{\underline{z}\underline{z}}\,,
\end{equation}

\noindent
and its effect on the 4-dimensional fields is a rescaling 

\begin{subequations}
  \begin{align}
    \label{eq:scalingtrafo1}
    \delta_{\hat{\eta}} k
    & =
      -k\,,
    &
      k' & = e^{-\alpha} k\,,
    \\
    & \nonumber \\
    \label{eq:scalingtrafo2}
    \delta_{\hat{\eta}}A_{\mu}
    & =
      A_{\mu}\,,
    &
      A' & = e^{\alpha} A\,,
    \\
    & \nonumber \\
    \delta_{\hat{\eta}}g_{\mu\nu}
    & =
      0\,,
    &
      g_{\mu\nu}' & = g_{\mu\nu}\,.
  \end{align}
\end{subequations}

Observe that the KK scalar will in general reach some constant value
$k_{\infty}\neq 1$ at spatial infinity. According to
Eq.~(\ref{eq:KKradius}), that value is related to the asymptotic value of the
radius $R$ and the length scale $\ell$ by

\begin{equation}
k_{\infty}= R/\ell\,.  
\end{equation}

\noindent Under the above rescaling

\begin{equation}
  k_{\infty}'
  =
  R/\ell^{'}
  =
  e^{-\alpha}k_{\infty}\,.
\end{equation}

Since the 5-dimensional theory we are starting with is diff-invariant, the
 4-dimensional one will also be invariant under these global
 rescalings. Observe that the vector field that generates these rescalings
 does not commute with the Killing vector that generates translations in the
 internal dimension

 \begin{equation}
   [\hat{\eta},\partial_{\underline{z}}]
   =
   -\partial_{\underline{z}}\,.
 \end{equation}

 Thus it does not commute with the vectors that generate 4-dimensional gauge
 transformations

 \begin{equation}
    [\hat{\eta},-\chi \partial_{\underline{z}}]
   =
   \chi\partial_{\underline{z}}\,,
   \,\,\,\,\,
   \Rightarrow
   \,\,\,\,\,
   [\delta_{\hat{\eta}},\delta_{\chi}]
   =
   -\delta_{\chi}\,.
 \end{equation}

Following Scherk and Schwarz \cite{Scherk:1979zr}, in order to find the
equations of motion that govern the dynamics of the 4-dimensional fields it is
convenient to use the Vielbein formalism, making a particular choice for the
decomposition of the 5-dimensional one $\hat{e}{}^{\hat{a}}{}_{\hat{\mu}}$ in
terms of the 4-dimensional fields $e^{a}{}_{\mu},A_{\mu},k$ that breaks the
5-dimensional Lorentz group down to the 4-dimensional one:

\begin{equation}
\label{eq:standardVielbeinAnsatz}
\left( \hat{e}^{\hat{a}}{}_{\hat{\mu}} \right) = 
\left(
\begin{array}{c@{\quad}c}
e^{a}{}_{\mu} & kA_{\mu} \\
&\\[-3pt]
0       & k    \\
\end{array}
\right)\!, 
\hspace{1cm}
\left(\hat{e}_{\hat{a}}{}^{\hat{\mu}} \right) =
\left(
\begin{array}{c@{\quad}c}
e_{a}{}^{\mu} & 0  \\
& \\[-3pt]
-A_{a}       & k^{-1} \\
\end{array}
\right)\,.
\end{equation}

\noindent
Here $A_{a}= e_{a}{}^{\mu} A_{\mu}$ and we will assume that all 4-dimensional
fields with Lorentz indices $a,b,c,\ldots$ have been contracted with the
4-dimensional Vielbein. The above expressions can also be written in the form

\begin{subequations}
  \begin{align}
    \hat{e}^{a}
    & =
      e^{a}\,, \hspace{1cm}
    &
      \hat{e}_{a}
    & =
      e_{a}-\imath_{a}A \partial_{\underline{z}}\,,
    \\
    & & & \nonumber \\
    \hat{e}^{z}
    & =
      k(dz+A)\,,
    &
      \hat{e}_{z}
    & =
      k^{-1}\partial_{\underline{z}}\,,
  \end{align}
\end{subequations}

\noindent
where $\imath_{\xi}$ indicates the inner product with the 4-dimensional vector
$\xi$ and $\imath_{a}$ with $e_{a}$, that is,
$\imath_{a}A = e_{a}{}^{\mu}A_{\mu}$.

With this decomposition, the non-vanishing components of the spin
connection\footnote{Our spin connection satisfies
  $\mathcal{D}e^{a}=de^{a}-\omega^{a}{}_{b}\wedge e^{b}=0$ in 5 and 4
  dimensions.}  are

\begin{equation}
\label{eq:standardspinconnectionreduction}
\begin{array}{rclrcl}
\hat{\omega}_{abc} & = & \omega_{abc}\,, &
\hat{\omega}_{abz} & = & \frac{1}{2} k F_{ab}\,,
\\
& & & & & \\
\hat{\omega}_{zbc} & = & -\frac{1}{2} k F_{bc}\,,
\hspace{1.5cm}& 
\hat{\omega}_{zbz} & = & -\partial_{b} \ln{k}\,,\\
\end{array}
\end{equation}

\noindent
or, written as 1-forms

\begin{equation}
  \begin{aligned}
    \hat{\omega}_{bc}
    & =
      \omega_{bc} -\tfrac{1}{2} k^{2} F_{bc}(dz+A)\,,
      \\
      & \\
      \hat{\omega}_{bz}
      & =
 \tfrac{1}{2} k F_{ab} e^{a}
      -\partial_{b}k\, (dz+A)\,.
  \end{aligned}
\end{equation}

The components of the curvature 2-form are\footnote{The curvature 2-form is
  defined by
  \begin{equation}
   R^{ab}= d\omega^{ab}-\omega^{a}{}_{c}\wedge \omega^{cb}\,, 
  \end{equation}
  both in 5 and in 4 dimensions.
}

\begin{equation}
  \begin{aligned}
    \hat{R}_{ab}
    & =
    R_{ab}
    -\tfrac{1}{2} k^{2} \left(F_{ab}F
      +\tfrac{1}{2}\imath_{a}F\wedge\imath_{b}F\right)
    -\tfrac{1}{2} \left[D\left(k^{2} F_{ab}\right)
    -\imath_{[a}dk^{2} \imath_{b]}F\right]\wedge (dz+A)\,.
  \\
  & \\
  \hat{R}_{az}
    & =
    -\tfrac{1}{2}\mathcal{D}\left( k \imath_{a}F\right) 
      -\imath_{a}dk F
    -\left[\mathcal{D}\imath_{a}dk\,
      -\tfrac{1}{4}k^{3}F_{a}{}^{c}\imath_{c}F\right]\wedge (dz+A)\,.
  \end{aligned}
\end{equation}

The Ricci 1-form
$\hat{R}^{\hat{c}}=\hat{R}_{\hat{\mu}}{}^{\hat{c}}dx^{\hat{\mu}}$ is defined
by

\begin{equation}
\hat{R}^{\hat{c}} \equiv\hat{\imath}_{\hat{b}}\hat{R}^{\hat{b}\hat{c}}\,,  
\end{equation}

\noindent
where $\hat{\imath}_{\hat{b}}$ is the inner product with the vector
5-dimensional vector
$\hat{e}_{\hat{b}} = \hat{e}_{\hat{b}}{}^{\hat{\mu}}\partial_{\hat{\mu}}$. We
have

\begin{equation}
  \begin{aligned}
    \hat{R}^{a}
    & =
    R^{a}+k^{-1}\mathcal{D}\imath^{a}dk +\tfrac{1}{2}k^{2}F^{ab}\imath_{b}F
    -\tfrac{1}{2}k^{-1}\mathcal{D}_{b}\left(k^{3} F^{ba}\right) (dz+A)\,,
    \\
    & \\
    \hat{R}^{z}
    & =
    \tfrac{1}{2}k^{-2}\mathcal{D}_{a}\left(k^{3}\imath^{a}F\right)
    +\left[\mathcal{D}^{2}k\,
      +\tfrac{1}{4}k^{3}F^{2} \right](dz+A)\,,
  \end{aligned}
\end{equation}

\noindent
where $\imath^{a} =\eta^{ab}\imath_{b}$.

Thus, the 5-dimensional Einstein equations $\hat{R}^{\hat{c}}=0$ are
equivalent to the following 3 equations involving 4-dimensional fields:

\begin{subequations}
\label{eq:4deomKKframe}
  \begin{align}
    R^{a}+k^{-1}\mathcal{D}\imath^{a}dk +\tfrac{1}{2}k^{2}F^{ab}\imath_{b}F
    & =
      0\,,
    \\
    & \nonumber \\
    \mathcal{D}_{b}\left(k^{3} F^{ba}\right)
    & =
      0\,,
      \\
    & \nonumber \\
    \mathcal{D}^{2}k\,
    +\tfrac{1}{4}k^{3}F^{2}
    & =
      0\,.
  \end{align}
\end{subequations}

The action from which these 4-dimensional equations can be derived can be
obtained from the 5-dimensional Einstein-Hilbert action
Eq.~(\ref{eq:5dEHaction}), which, in the Vielbein formalism, takes the form

\begin{equation}
  \label{eq:5dEHactionVielbein}
  S[\hat{e}]
  =
  \frac{1}{16\pi G_{N}^{(5)}}
  \int \hat{\star}(\hat{e}^{\hat{a}}\wedge \hat{e}^{\hat{b}})
  \wedge \hat{R}_{\hat{a}\hat{b}}\,.
\end{equation}

Substituting the above decompositions of the 5-dimensional Vielbein and
curvature in terms of the 4-dimensional fields we get

\begin{equation}
  \label{eq:4dEHactionVielbein0}
  S[e,A,k]
  =
  \frac{1}{16\pi G_{N}^{(5)}}
  \int \left\{ k\left[ -\star(e^{a}\wedge e^{b})
      \wedge R_{ab} +\tfrac{1}{2}k^{2}F\wedge \star F \right]
    +d\left[2\star dk\right]\right\}\wedge dz\,.
\end{equation}

\noindent
Integrating over the internal coordinate $z\in [0,2\pi \ell]$ and using the
$z$-dependence of the 4-form, we get

\begin{equation}
  \label{eq:4dEHactionVielbein}
  S[e,A,k]
  =
  \frac{2\pi\ell}{16\pi G_{N}^{(5)}}
  \int \left\{ k\left[ -\star(e^{a}\wedge e^{b})
      \wedge R_{ab} +\tfrac{1}{2}k^{2}F\wedge \star F \right]
    +d\left[2\star dk\right]\right\}\,.
\end{equation}

We have kept the total derivative because total derivatives can modify Noether
currents and charges.

It is not too difficult to see that the equations that one gets from this
action are combinations of Eqs.~(\ref{eq:4deomKKframe}) and, therefore,
equivalent to them.

The factor of $k$ in front of the Einstein-Hilbert term in
Eq.~(\ref{eq:4dEHactionVielbein0}) indicates that the 4-dimensional metric
$g_{\mu\nu}$ is not in the (conformal) Einstein frame, in which, by
definition, the Einstein-Hilbert term has no additional scalar factors. The
Einstein-frame metric is clearly related to $g_{\mu\nu}$ by a Weyl rescaling
with some power of the KK scalar $k$. If we want the rescaling to preserve the
normalization of the metric in the non-compact directions, we must use a power
of $k/k_{\infty}$ and not just of $k$ to rescale it. Thus, we define the
4-dimensional Einstein-frame metric $g_{E\, \mu\nu}$ and Vielbein
$e_{E}{}^{a}{}_{\mu}$ and the Einstein-frame KK vector field $A_{E\, \mu}$ by

\begin{equation}
  g_{\mu\nu}= \left(k/k_{\infty}\right)^{-1}g_{E\, \mu\nu}\,,
  \hspace{.5cm}
  e^{a}{}_{\mu} = \left(k/k_{\infty}\right)^{-1/2}e_{E}{}^{a}{}_{\mu}\,,
  \hspace{.5cm}
  A_{\mu} = k_{\infty}^{1/2}A_{E\, \mu}\,.
\end{equation}

Under this rescaling,

\begin{equation}
  \begin{aligned}
    R^{ab}
    & =
      R_{E}^{ab}
      -\mathcal{D}_{E}
      \left(\imath_{E}^{[a}d\log{k}e_{E}{}^{b]}\right)
      -\tfrac{1}{2}\imath_{E}^{[a}d\log{k}
      d\log{k} \wedge e_{E}{}^{b]}
      \\
    &  \\
    & \hspace{.5cm}
      +\tfrac{1}{4} \left(\partial\log{k}\right)^{2}
      e_{E}{}^{a}\wedge e_{E}{}^{b}\,,
  \end{aligned}
\end{equation}

\noindent
and

\begin{equation}
  \begin{aligned}
    -k\star(e^{a}\wedge e^{b}) \wedge R_{ab}
    & = k_{\infty} \left\{
      -\star_{E}(e_{E}{}^{a}\wedge e_{E}{}^{b}) \wedge R_{E\, ab}
      +\tfrac{3}{2}d\log{k}\wedge \star_{E} d\log{k}
    \right.
    \\
    & \\
    & \hspace{.5cm}
    \left.
      +d\left[-3\star_{E}d\log{k}\right] \right\}\,.
  \end{aligned}
\end{equation}

\noindent
We arrive to the Einstein-frame action 

\begin{equation}
  \label{eq:4dEHactionVielbeinEframe}
  \begin{aligned}
    S[e_{E},A_{E},k]
    & =
      \frac{1}{16\pi G_{N}^{(4)}}
      \int \left\{ -\star_{E}(e_{E}{}^{a}\wedge e_{E}{}^{b})
      \wedge R_{E\, ab}
      +\tfrac{3}{2}d\log{k}\wedge \star_{E} d\log{k}
       \right.
    \\
    & \\
    & \hspace{.5cm}
    \left.
      +\tfrac{1}{2}k^{3}F_{E}\wedge \star F_{E}
      +d\left[-\star_{E}d\log{k}\right]\right\}\,,
      \\
  \end{aligned}
\end{equation}

\noindent
with the 4-dimensional Newton constant given by 

\begin{equation}
  \label{eq:4-5Newtonconstant}
  G_{N}^{(4)}
  =
  \frac{G_{N}^{(5)}}{2\pi R}\,.
\end{equation}

\noindent
Observe that the 4-dimensional Newton constant depends on the invariant radius
$R$ and not on $\ell$ or $k_{\infty}$, both of which transform under
rescalings.

Finally, we redefine $k$ in terms of an unconstrained scalar field $\phi$
which can take any real value

\begin{equation}
  k
  =
  e^{\phi/\sqrt{3}}\,,
\end{equation}

\noindent
and we arrive to the final form of our action

\begin{equation}
  \label{eq:4dEHactionVielbeinEframestandard}
  \begin{aligned}
    S[e_{E},A_{E},\phi]
    & =
      \frac{1}{16\pi G_{N}^{(4)}}
      \int \left\{ -\star_{E}(e_{E}{}^{a}\wedge e_{E}{}^{b})
      \wedge R_{E\, ab}
      +\tfrac{1}{2}d\phi\wedge \star_{E} d\phi
      +\tfrac{1}{2}e^{\sqrt{3}\phi}F_{E}\wedge \star_{E} F_{E} \right\}
    \\
    & \\
    & \hspace{.5cm}
      +\frac{1}{16\pi G_{N}^{(4)}}
      \int d\left( -\tfrac{1}{\sqrt{3}}\star_{E}d\phi\right)\,.
  \end{aligned}
\end{equation}

This is a particular Einstein-Maxwell-dilaton (EMD) model with $a=-\sqrt{3}$
in the parametrization used in Ref.~\cite{Ballesteros:2023iqb}, whose results
we can use as long as we take into account the additional total derivative
term.\footnote{See also Ref.~\cite{Pacilio:2018gom} for the definitionof the
  scalar charge.}

After all these redefinitions, the relation between the 5-dimensional line
element and the 4-dimensional Einstein-frame line element

\begin{equation}
  ds^{2}_{E\,(4)}
  =
  g_{E\, \mu\nu}dx^{\mu}dx^{\nu}\,,
\end{equation}

\noindent
and other Einstein-frame fields is

\begin{equation}
  \label{eq:5dimensionalmetric}
  ds_{(5)}^{2}
  =
  e^{-(\phi-\phi_{\infty})/\sqrt{3}}ds^{2}_{E\,(4)}
  - e^{2\phi/\sqrt{3}}
  \left[dz+e^{\frac{\phi_{\infty}}{2\sqrt{3}}}A_{E}\right]^{2}\,,
\end{equation}

\noindent
where

\begin{equation}
  e^{\phi_{\infty}/\sqrt{3}}
  =
  k_{\infty}\,.
\end{equation}

Given a solution of the 4-dimensional theory
Eq.~(\ref{eq:4dEHactionVielbeinEframestandard}) with fields
$g_{E\,\mu\nu},A_{E\,\mu},\phi$, with $\phi \rightarrow \phi_{\infty}$ at
infinity, the above relation allows us to rewrite it as a solution of pure
5-dimensional gravity. In the next section we are going to review the static,
electrically charged, spherically symmetric 4-dimensional black hole solutions
of this theory.

Let us now derive the equations of motion of the 4-dimensional theory.  Under
a general variation of the fields, the action
Eq.~(\ref{eq:4dEHactionVielbeinEframestandard}) behaves as follows:

\begin{equation}
\delta  S[e_{E},A_{E},k_{E}]
  =
   \int  \left\{\mathbf{E}_{E\, a}\wedge \delta e_{E}{}^{a}
    +\mathbf{E}_{E\, \phi}\delta \phi 
    +\mathbf{E}_{E\, A_{E}}\delta A_{E}
    +\mathbf{\Theta}(\varphi,\delta\varphi)\,,
    \right\}
\end{equation}

\noindent
where $\varphi$ denotes collectively all the fields of the theory. 
Suppressing the overall factors of $(16\pi G_{N}^{(4)})^{-1}$, the equations
of motion are given by 

\begin{subequations}
  \begin{align}
    \label{eq:Ea}
    \mathbf{E}_{E\, a}
    & =
      \imath_{E\, a}\star_{E}(e_{E}^{a}\wedge e_{E}^{b})\wedge R_{E\, ab}
      +\tfrac{1}{2}\left(\imath_{E\, a}d\phi\wedge \star_{E} d\phi
      +d\phi\wedge \imath_{E\, a}\star_{E} d\phi \right)
      \nonumber \\
    & \nonumber \\
    & \hspace{.5cm}
      +\tfrac{1}{2}e^{\sqrt{3}\phi}\left(\imath_{E\, a}F_{E}\wedge \star_{E} F_{E}
      -F_{E}\wedge \imath_{E\, a}\star_{E} F_{E} \right)\,,
    \\
    & \nonumber \\
    \label{eq:Ephi}
    \mathbf{E}_{E\, \phi}
    & =
      -d(\star_{E}d\phi)
      +\tfrac{\sqrt{3}}{2} e^{\sqrt{3}\phi}F_{E}\wedge \star _{E}F_{E}\,, 
    \\
    & \nonumber \\
        \label{eq:EA}
    \mathbf{E}_{E\, A_{E}}
    & = -d\left(e^{\sqrt{3}\phi}\star_{E}F_{E}\right)\,,
  \end{align}
\end{subequations}

\noindent
while

\begin{equation}
  \label{eq:Theta}
  \begin{aligned}
    \mathbf{\Theta}(\varphi,\delta \varphi)
    & =
    -\star_{E}(e_{E}^{a}\wedge
    e_{E}^{b})\wedge \delta \omega_{E\, ab} +e^{\sqrt{3}\phi} \star
    _{E}F_{E}\wedge\delta A_{E}
    +\star_{E}d\phi\delta\phi
    \\
    & \\
    & \hspace{.5cm}
    -\tfrac{1}{\sqrt{3}}\left[\star_{E}d\delta \phi
      +\imath_{E\, a}\star_{E}d\phi\wedge \delta e_{E}{}^{a}
    -\star_{E}\delta e_{E}{}^{a}\imath_{E\, a}d\phi\right]\,.
  \end{aligned}
\end{equation}

The action and equations of motion are evidently invariant under U$(1)$ gauge
transformations

\begin{equation}
\label{eq:vectorgaugetransformations}
  \delta_{\chi_{E}}A_{E} = d\chi_{E}\,,
  \hspace{1cm}
  \chi_{E} \equiv k_{\infty}^{1/2}\chi\,,
\end{equation}

\noindent
as well as under the global scale transformations

\begin{equation}
  \label{eq:globalsym}
  \delta_{\alpha}\phi = \sqrt{3}\alpha\,,
  \hspace{1cm}
  \delta_{\alpha}A_{E} = e^{-3\alpha/2}A_{E}\,,
\end{equation}

\noindent
which originate in the 5-dimensional general coordinate transformation
$z'=e^{\alpha}z$.

The set of equations of motion of the 4-dimensional KK theory, enhanced with
the Bianchi identity of the KK vector field strength

\begin{equation}
\mathbf{B}_{E} \equiv dF_{E}=0\,,  
\end{equation}

\noindent
are left invariant by the electric-magnetic duality transformation of the
vector and scalar fields

\begin{equation}
  \label{eq:emdualitytransformations}
  F_{E}'= e^{-\sqrt{3}\phi}\star F_{E}\,,
  \hspace{1cm}
  \phi' = -\phi\,.
\end{equation}

This transformation can be used to generate new from already known solutions
and, in particular, magnetic from electric solutions and vice versa. In
contrast with the other symmetries of the equations of motion of the KK theory
(diffeomorphisms, U$(1)$ gauge transformations and global rescalings) it is
not clear whether this symmetry has a 5-dimensional, purely geometrical
origin. To start with, it is unclear what the 5-dimensional origin of the
Bianchi identity $\mathbf{B}_{E}=0$ is. On the other hand, this is a symmetry
of the equations of motion only: the transformations
Eqs.~(\ref{eq:emdualitytransformations}) do not leave invariant the action.

Finally, notice that the scalar equation can be rewritten in the alternative
form

\begin{equation}
      \label{eq:Ephialt}
    \mathbf{E}_{E\, \phi}
    =
      -d\left[\star_{E}d\phi 
        -\tfrac{\sqrt{3}}{2} e^{\sqrt{3}\phi}A_{E}\wedge \star _{E}F_{E}\right]
      -\tfrac{\sqrt{3}}{2} A_{E}\wedge \mathbf{E}_{A}\,.
\end{equation}

The term in square brackets is the on-shell conserved Noether $(d-1)$-form
current associated to the invariance under the global transformations
Eq.~(\ref{eq:globalsym}).

\subsection{Motion in a Kaluza-Klein spacetime}
\label{eq:motioninKKspacetime}

It is interesting to study the geodesic motion of test particles in the
5-dimensional space, which is controlled by the equations

\begin{subequations}
  \label{eq:d5geodesicequations}
  \begin{align}
    \ddot{x}^{\hat{\mu}}
    +\hat{\Gamma}_{\hat{\nu}\hat{\rho}}{}^{\hat{\mu}}\dot{x}^{\hat{\nu}}
    \dot{x}^{\hat{\rho}}
    & =
      0\,,
    \\
    & \nonumber \\
    \hat{g}_{\hat{\mu}\hat{\nu}}\dot{x}^{\hat{\mu}}\dot{x}^{\hat{\nu}}
    & =
      \alpha\,,
  \end{align}
\end{subequations}

\noindent
where $\alpha=0$ for massless particles and $\alpha=m^{2}$ for massive
particles.

Rewriting these equations in terms of the 4-dimensional fields, we find

\begin{subequations}
  \label{eq:d4geodesicequations}
  \begin{align}
    \ddot{x}^{\mu}
    +\Gamma_{\nu\rho}{}^{\mu}\dot{x}^{\nu} \dot{x}^{\rho}
    -F_{\nu}{}^{\mu}\dot{x}^{\nu} k^{2}(\dot{z}+A_{\rho}\dot{x}^{\rho})
    -\tfrac{1}{2}\partial^{\mu}k^{2}(\dot{z}+A_{\rho}\dot{x}^{\rho})^{2}
    & =
      0\,,
    \\
    & \nonumber \\
    g_{\mu\nu}\dot{x}^{\mu}\dot{x}^{\nu} -k^{2}(\dot{z}+A_{\rho}\dot{x}^{\rho})^{2}
    & =
      \alpha\,,
  \end{align}
\end{subequations}

\noindent
plus the equation for $z(\xi)$. This equation is complicated but it is
entirely equivalent to the conservation of the momentum conjugate to $z$

\begin{equation}
  P_{z}
  =
  -k^{2}(\dot{z}+A_{\rho}\dot{x}^{\rho})\,.
\end{equation}

\noindent
Using this relation to eliminate $\dot{z}$ in
Eqs.~(\ref{eq:d4geodesicequations}), they take the form

\begin{subequations}
  \label{eq:d4geodesicequations2}
  \begin{align}
    \ddot{x}^{\mu}
    +\Gamma_{\nu\rho}{}^{\mu}\dot{x}^{\nu} \dot{x}^{\rho}
    & =
   P_{z}F^{\mu}{}_{\nu}\dot{x}^{\nu} 
    -\tfrac{1}{2}P_{z}^{2}\partial^{\mu}k^{-2}\,,
    \\
    & \nonumber \\
    g_{\mu\nu}\dot{x}^{\mu}\dot{x}^{\nu} 
    & =
      \alpha +k^{-2}P_{z}^{2}\,,
  \end{align}
\end{subequations}

\noindent
which are the equations of motion of a 4-dimensional particle with electric
charge $P_{z}$ and a spacetime-dependent effective mass squared
$\alpha+k^{-2}P_{z}^{2}$. The interaction with the scalar induces another
force term proportional to $P_{z}^{2}$.

Eqs.~(\ref{eq:d5geodesicequations}) can be derived from the Polyakov-type
action

\begin{equation}
  \label{eq:d5Polyakov}
  \hat{S}[e,x^{\hat{\mu}}]
  =
  -\tfrac{1}{2}\int d\xi
  \left\{e^{-1}\hat{g}_{\hat{\mu}\hat{\nu}}\dot{x}^{\hat{\mu}}\dot{x}^{\hat{\nu}}+e
  m^{2}\right\}\,.
\end{equation}

When $m\neq 0$, $e$ can be eliminated replacing the (algebraic) solution to its
equation of motion in the above action. One obtains the Nambu-Goto-type action

\begin{equation}
  \label{eq:d5Nambu-Goto}
  \hat{S}[x^{\hat{\mu}}]
  =
  -m\int d\xi\,
  \sqrt{\hat{g}_{\hat{\mu}\hat{\nu}}\dot{x}^{\hat{\mu}}\dot{x}^{\hat{\nu}}}\,.
\end{equation}

Rewriting  the action Eq.~(\ref{eq:d5Polyakov}) in terms of the 4-dimensional
fields and performing a Legendre transformation to eliminate $z(\xi)$, one
arrives to 

\begin{equation}
  \label{eq:d4Polyakov}
  \hat{S}[e,x^{\mu}]
  =
  -\tfrac{1}{2}\int d\xi
  \left\{e^{-1}g_{\mu\nu}\dot{x}^{\mu}\dot{x}^{\nu}+e
  \left[m^{2}+k^{-2}P_{z}^{2}\right] +2P_{z} A_{\rho}\dot{x}^{\rho}\right\}\,,
\end{equation}

\noindent
and eliminating $e$ and using the fact that $P_{z}$ is constant, we get

\begin{equation}
  \label{eq:d4Nambu-Goto}
  \hat{S}[x^{\mu}]
  =
  -\int d\xi \sqrt{m^{2}+k^{-2}P_{z}^{2}}
  \sqrt{g_{\mu\nu}\dot{x}^{\mu}\dot{x}^{\nu}}
  -P_{z}\int d\xi A_{\rho}\dot{x}^{\rho}\,.
\end{equation}

The physical interpretation of this action is exactly the same as that of
Eqs.~(\ref{eq:d4geodesicequations2}), which, as expected, can be derived from
Eq.~(\ref{eq:d4Polyakov}).

In the Einstein frame, the above action takes the form

\begin{equation}
  \label{eq:d4Nambu-Goto-Einsteinframe}
  \hat{S}[x^{\mu}]
  =
  -\int d\xi \sqrt{m^{2}(k/k_{\infty})^{-1} +k^{-3}q^{2}}
  \sqrt{g_{E\, \mu\nu}\dot{x}^{\mu}\dot{x}^{\nu}}
  -q\int d\xi A_{E\, \rho}\dot{x}^{\rho}\,,
\end{equation}

\noindent
and it describes a particle of electric charge

\begin{equation}
  \label{eq:chargeversusmomentum}
  q
  =
  P_{z}k_{\infty}^{1/2}\,,
\end{equation}

\noindent
and a position-dependent inertial mass that depends on the 5-dimensional mass
and the charge and their couplings to the KK scalar.

\section{The electric  Kaluza-Klein black hole}
\label{sec-electricKKBH}

The basic static, spherically symmetric, asymptotically flat, electric,
solution of the 4-dimensional KK theory
Eq.~(\ref{eq:4dEHactionVielbeinEframestandard}) has the form \footnote{As we
  have mentioned before, the theory at hand is the $a=-\sqrt{3}$ EMD
  theory. The static spherically symmetric solutions of these theories for all
  values of $a$ were found in
  Refs.~\cite{Gibbons:1982ih,Gibbons:1984kp,Holzhey:1991bx}. Here we use the
  form in which it is written in Ref.~\cite{Ortin:2015hya}.}

\begin{equation}
  \label{eq:4delectricsolution}
  \begin{aligned}
    ds_{E\, (4)}^{2}
    & =
    H^{-1/2}Wdt^{2}-H^{1/2}\left(W^{-1}dr^{2}+r^{2}d\Omega_{(2)}^{2}\right)\,,
    \\
    & \\
    A_{E}
    & =
    \alpha  e^{-\sqrt{3}\phi_{\infty}/2}\left(H^{-1}-1\right)dt\,,
    \\
    & \\
    e^{\sqrt{3}\phi}
    & =
    e^{\sqrt{3}\phi_{\infty}}H^{3/2}\,,
  \end{aligned}
\end{equation}

\noindent
where the functions $H$ and $W$ are given by

\begin{equation}
\label{eq:HandWfunctions}
  H
  =
  1+\frac{h}{r}\,,
  \hspace{1cm}
  W
  =
  1+\frac{w}{r}\,,
\end{equation}

\noindent
and the integration constants $h,w,\alpha$ satisfy the relation

\begin{equation}
\label{eq:relation}
w = h(1-\alpha^{2})\,.  
\end{equation}

In this solution we can always take $h\geq 0$ and $w\leq 0$. Therefore, for
$w\neq 0$, there is always a regular horizon at $r=r_{0}=-w$. When $w=0$
($r_{0}=0$), the horizon becomes singular and the solution is not a black
hole. Multicenter solutions are possible in this case and we will refer to it
as the extremal case and to $r_{0}$ as the non-extremality parameter.

Defining the mass $M$ through the asymptotic expansion

\begin{equation}
g_{tt} \sim 1-\frac{2G_{N}^{(4)}M}{r} +\mathcal{O}\left(\frac{1}{r^{2}}\right)\,,
\end{equation}

\noindent
and the electric charge through the integral\footnote{Notice that, with this
  definition, the electric charge $q$ rescales as
  \begin{equation}
  q \rightarrow e^{-\alpha/2}q\,,  
  \end{equation}
  aunder the transformations Eqs.~(\ref{eq:scalingtrafo1}) and
  (\ref{eq:scalingtrafo2}).}

\begin{equation}
    \label{eq:electricchargedef}
    q
     =
      \frac{1}{16\pi G_{N}^{(4)}}\int_{\Sigma^{2}}e^{\sqrt{3}\phi} \star_{E}
      F_{E}\,,
\end{equation}

\noindent
where $\Sigma^{2}$ is a 2-dimensional closed surface that encloses the black
hole\footnote{Usually, it is taken to be the 2-sphere at spatial infinity, but
  one gets the same result integrating on any other surface that can be
  obtained by deforming it, such as sections of the horizon, as long as we do
  not cross any singularities, the equations of motion are satisfied and the
  integrands remain closed 2-forms.} we find that the integration constants
$w,h,\alpha$ can be expressed in terms of $M$ and $q$ as follows:

\begin{subequations}
  \label{eq:whandphysicalparameters}
  \begin{align}
    -w
    & =
    r_{0}
     =
      G_{N}^{(4)}\left[3M -\sqrt{M^{2}+8e^{-\sqrt{3}\phi_{\infty}}q^{2}}\right]\,,
    \\
    & \nonumber \\
    h
    & =
      -2G_{N}^{(4)}\left[M -\sqrt{M^{2}+8e^{-\sqrt{3}\phi_{\infty}}q^{2}}\right]\,,
    \\
    & \nonumber \\
\alpha
    & =
      \frac{2e^{-\sqrt{3}\phi_{\infty}/2}q}{M -\sqrt{M^{2}+8e^{-\sqrt{3}\phi_{\infty}}q^{2}} }\,.
  \end{align}
\end{subequations}

Then, extremality is reached when

\begin{equation}
  \label{eq:extremalityconditionelectric}
  M = e^{-\sqrt{3}\phi_{\infty}/2}|q|\,,
\end{equation}

One can also define a charge for the scalar field, $\Sigma$, through its
asymptotic expansion at spatial infinity

\begin{equation}
  \phi
  \sim
  \phi_{\infty} +\frac{G_{N}^{(4)}\Sigma}{r}
  +\mathcal{O}\left(\frac{1}{r^{2}}\right)\,.  
\end{equation}

\noindent
or equivalently, through the definition proposed in
Ref.~\cite{Pacilio:2018gom,Ballesteros:2023iqb}.

The scalar charge $\Sigma$ is not expected to be an independent quantity
(``charge'') characterizing a black hole (\textit{i.e.}~an actual black hole
with regular event horizon). As a matter of fact, the solutions of this theory
in which $\Sigma$ is independent do not have an event horizon (they are
singular) and in the solutions at hand  do have a regular event horizon
$\Sigma$ is given by the following function of the conserved charges
\cite{Ballesteros:2023iqb}:

\begin{equation}
  \label{eq:SigmaKKelectric}
  \Sigma
  =
  -\sqrt{3}\left[M-\sqrt{M^{2}+8e^{-\sqrt{3}\phi_{\infty}}q^{2}}\right]\,.
\end{equation}

Let us now consider the thermodynamical properties of this black hole. This
black hole, and the rest of the black holes that we are going to consider,
being static, have a timelike Killing vector, $l \equiv \partial_{t}$ and
their event horizons, located at $r=r_{0}$ when $r_{0}\neq 0$, coincide with
the associated Killing horizon

\begin{equation}
  l^{2}
  =
  g_{tt}
  =
  0\,.
\end{equation}

The Hawking temperature and the Bekenstein-Hawking entropy are given by

\begin{subequations}
  \begin{align}
    \label{eq:temperatureKK}
    T
    & =
    \frac{1}{4\pi[r_{0}(r_{0}+h)]^{1/2}}\,,
    \\
    & \nonumber \\
    \label{eq:entropyKK}
    S
    & =
    \frac{\pi}{G_{N}^{(4)}}\left(r_{0}+h\right)^{1/2}r_{0}^{3/2}\,,    
  \end{align}
\end{subequations}

\noindent
and they satisfy the relation

\begin{equation}
  \label{eq:2STrelation}
  2ST
  =
  \frac{r_{0}}{2G_{N}^{(4)}}\,.
\end{equation}

Observe that, in the extremal limit $r_{0}\rightarrow 0$, the temperature
goes to infinity while the entropy goes to zero.

Eq.~(\ref{eq:2STrelation}) can be used to derive a Smarr-like formula
multiplying and dividing the right-hand side by $r_{0}+h$. In the first
case, we get

\begin{equation}
  M
  =
  2ST
  +\frac{4 e^{-\sqrt{3}\phi_{\infty}}q}{M+\sqrt{M^{2}+8e^{-\sqrt{3}\phi_{\infty}}q^{2}} }q\,,
\end{equation}

\noindent
and, comparing with

\begin{equation}
  \label{eq:Smarr}
  M
  =
  2ST +\Phi q\,,  
\end{equation}

\noindent
we can identify

\begin{equation}
  \label{eq:PhiKK}
  \Phi
  =
  \frac{4 e^{-\sqrt{3}\phi_{\infty}}q}{M+\sqrt{M^{2}+8e^{-\sqrt{3}\phi_{\infty}}q^{2}} }\,,  
\end{equation}

\noindent
with the difference between the value of electrostatic potential at the
horizon and at spatial infinity. Indeed, one can check that, in the gauge we
are using, in which the potential vanishes at spatial infinity

\begin{equation}
  \label{eq:Phidef}
\Phi = A_{E,t}(r_{0})\,.
\end{equation}

Varying the entropy Eq.~(\ref{eq:entropyKK}) we get

\begin{equation}
  \label{eq:firstlaw}
  \delta S
  =
  \frac{1}{T}\left(\delta M -\Phi\delta q
    -\tfrac{1}{4}\Sigma\delta\phi_{\infty}\right)\,,
\end{equation}

\noindent
where $T,\Phi,\Sigma$ are given, respectively, by
Eqs.~(\ref{eq:temperatureKK}), (\ref{eq:PhiKK}) and (\ref{eq:SigmaKKelectric}),
confirming our identification of the thermodynamical potentials conjugated to
the charges.





\section{4-dimensional thermodynamics \textit{\`a la Wald}}
\label{sec-4dthermodynamics}

As we have mentioned before, the 4-dimensional action
Eq.~(\ref{eq:4dEHactionVielbeinEframestandard}) is a particular example of the
Einstein-Maxwell-dilaton (EMD) model with $a=-\sqrt{3}$, up to a total
derivative. The thermodynamics of the black-hole solutions of all these models
has been studied in detail using Wald's formalism in
Ref.~\cite{Ballesteros:2023iqb}.  It can easily be shown that the total
derivative has no effect on the first law, which is given by
Eq.~(\ref{eq:firstlaw}) when the magnetic terms are set to zero.

For later use we quote the expression of the Noether-Wald charge

\begin{equation}
  \label{eq:4dNWcharge}
  \mathbf{Q}[\xi]
  =
  \frac{1}{16\pi G_{N}^{(4)}}
  \left\{\star_{E}(e_{E}{}^{a}\wedge e_{E}{}^{b})P_{E\,\xi\,ab}
    -e^{\sqrt{3}\phi}\star_{E}F_{E}P_{E\,\xi}
    +\tfrac{1}{\sqrt{3}}\imath_{\xi}\star_{E}d\phi
    \right\}\,.
\end{equation}

With this charge we can construct the generalized Komar charge
\cite{Komar:1958wp} by the procedure explained in
Refs.~\cite{Liberati:2015xcp,Ortin:2021ade,Mitsios:2021zrn}. First, we compute
$\omega_{l}$, defined by

\begin{equation}
  \label{eq:ilL}
\imath_{l}\mathbf{L}
\doteq
d\omega_{l}\,,
\end{equation}

\noindent
which in this case is given by

\begin{equation}
  \label{eq:omegal}
  \omega_{l}
  =
-\tfrac{1}{2}e^{\sqrt{3}\phi}\star_{E}F_{E}P_{E\,l}
  -\tfrac{1}{2}F_{E}\tilde{P}_{E\,l}
  +\tfrac{1}{\sqrt{3}}\imath_{l}\star_{E}d\phi\,.
\end{equation}

Then, the Komar charge is given by

\begin{equation}
  \label{eq:4dKomarcharge}
  \mathbf{K}[l]
  =
  \mathbf{Q}[l]-\omega_{l}
  =
    \frac{1}{16\pi G_{N}^{(4)}}
  \left\{\star_{E}(e_{E}{}^{a}\wedge e_{E}{}^{b})P_{E\,l\,ab}
    -\tfrac{1}{2}\left[e^{\sqrt{3}\phi}\star_{E}F_{E} P_{E\,l}
      -F_{E}\tilde{P}_{E\,l}\right]
    \right\}\,,
\end{equation}

\noindent
and, for purely electric black holes, it leads to the Smarr formula
Eq.~(\ref{eq:Smarr}).

Our goal now is to recover these results form the purely 5-dimensional point
of view. First, we are going to study some general aspects of the
5-dimensional geometry of 4-dimensional Kaluza-Klein black holes.

\section{Kaluza-Klein black-hole solutions from the 5-dimensional point of view}
\label{sec-KKBHs5}

An important previous question is whether the presence of event horizons in
the 4-dimensional metric implies their presence in the 5-dimensional one. We
need to study the behaviour of null geodesics in the 5-dimensional spacetime
which is determined by Eqs.~(\ref{eq:d5geodesicequations}) which we have shown
to be equivalent to the 4-dimensional Eqs.~(\ref{eq:d4geodesicequations2})
plus the equation of conservation of $P_{z}$.  The second of
Eqs.~(\ref{eq:d4geodesicequations2}) is particularly interesting because it
tells us that the lightcones of the 5-dimensional metric are equal to those of
the 4-dimensional one, times a circle.\footnote{The conformal rescaling that
  brings us to the Einstein metric leaves the lightcones invariant.}
5-dimensional, massless, $P_{z}\neq 0$ particles which move over the
5-dimensional lightcone are seen to move inside the 4-dimensional one. In
particular, this means that, if the 4-dimensional metric has event horizons,
so does the 5-dimensional one at the same place in the 4-dimensional
coordinates. The 5-dimensional horizon simply has one more dimension,
parametrized by $z$, fibered over the 4-dimensional one and we will denote
both the 4- and 5-dimensional event horizons by $\mathcal{H}$.

The main feature of the 5-dimensional geometries that correspond to the
4-dimensional static black holes we are considering is the fact that they all
admit the 5-dimensional spacelike Killing vector
$\hat{k}=\partial_{\underline{z}}$ and a timelike Killing vector associated to
the staticity of the 4-dimensional metric, $l=\partial_{t}$. The event horizon
of the 4-dimensional black hole is also the Killing horizon of this vector and
can be characterized by this property

\begin{equation}
l^{2}=l^{\mu}g_{\mu\nu}l^{\nu} \stackrel{\mathcal{H}}{=}0\,.  
\end{equation}

As we have just discussed, there is a 5-dimensional event horizon which is a
S$^{1}$ fibration over the 4-dimensional event horizon $\mathcal{H}$ and the
4-dimensional timelike Killing vector $l=\partial_{t}$ is also a Killing
vector of the 5-dimensional metric.  However, if we use the 5-dimensional
metric

\begin{equation}
  l^{\mu}\hat{g}_{\mu\nu}l^{\nu} = l^{2}-k^{2}(\imath_{l}A)^{2}
  \stackrel{\mathcal{H}}{=} -k^{2}(\imath_{l}A)^{2}\,.
\end{equation}

Thus, from the 5-dimensional point of view, the event horizon is not the
Killing horizon of $l$.

It is natural to search for a 5-dimensional extension of $l$, that we will
denote by $\hat{l}$, whose Killing horizon coincides with the event horizon,
that is,

\begin{equation}
  \hat{l}^{2}=\hat{l}^{\hat{\mu}}\hat{g}_{\hat{\mu}\hat{\nu}}\hat{l}^{\hat{\nu}}
  \stackrel{\mathcal{H}}{=}0\,.   
\end{equation}

\noindent
Assuming that $\hat{l}$ has the form

\begin{equation}
\hat{l}=l+f\hat{k}\,,  
\end{equation}

\noindent
we have

\begin{equation}
  \hat{l}^{2}
  =
  l^{\mu}\hat{g}_{\mu\nu}l^{\nu}
  -f^{2}\hat{g}_{\underline{z}\underline{z}}
  -2f  l^{\mu}\hat{g}_{\mu\underline{z}}
  =
  l^{2} -k^{2}(f+\imath_{l}A)^{2}
  \stackrel{\mathcal{H}}{=}
  -\left.k^{2}(f+\imath_{l}A)^{2}\right|_{\mathcal{H}}\,,     
\end{equation}

\noindent
which means that  we have to demand that

\begin{equation}
  \left.(f+\imath_{l}A)\right|_{\mathcal{H}}
=
0\,,
\end{equation}

\noindent
and we are going to assume that

\begin{equation}
  \label{eq:fg}
  f=-\imath_{l}A +\gamma\,,
  \,\,\,\,
  \text{where}
  \,\,\,\,
  \left. \gamma\right|_{\mathcal{H}}
  =
  0\,.
\end{equation}

\noindent
While this is not the most general possibility, it will be good enough for us.

On the other hand, we want $\hat{l}$ to be a Killing vector of the
5-dimensional metric. If $\partial_{\underline{z}}f=0$,

\begin{equation}
  \pounds_{\hat{l}}\hat{g}_{\underline{z}\underline{z}}
  =
  -2k\pounds_{l}k\,,
  \,\,\,\,\,
  \Rightarrow
  \pounds_{l}k=0\,,
\end{equation}

\noindent
which is satisfied in the solutions considered.  Furthermore, taking into
account the previous result,

\begin{equation}
  \pounds_{\hat{l}}\hat{g}_{\mu\underline{z}}
  =
  -2kA_{\mu}\pounds_{l}k -k^{2}\left(\pounds_{l}A_{\mu}+\partial_{\mu}f\right)\,,
  \,\,\,\,\,
  \Rightarrow
  \pounds_{l}A_{\mu}+\partial_{\mu}f=0\,,
\end{equation}

\noindent
and we conclude that the Lie derivative of the 1-form must vanish up to a
gauge transformation with parameter $f$. Using Eq.~(\ref{eq:fg}) and
differential-form language, and rescaling the equation with
$k^{-1/2}_{\infty}$, this condition takes the form

\begin{equation}
  \imath_{l}F_{E}+d(k^{-1/2}_{\infty}\gamma)=0\,.  
\end{equation}

\noindent
This is nothing but the Maxwell momentum map equation introduced in
Ref.~\cite{Elgood:2020svt} with $k^{-1/2}_{\infty}g$ playing the role of
momentum map $ P_{E\, l}$, and, taking into account that
$ \left. \gamma\right|_{\mathcal{H}} = 0$ and that the momentum map is defined
only up to an additive constant, we conclude that

\begin{equation}
  k^{-1/2}_{\infty}\gamma
  =
  P_{E\, l}-\left.P_{E\,l}\right|_{\mathcal{H}}
  \equiv
  \overline{P}_{E\, l}\,,
\end{equation}

\noindent
and we arrive at 

\begin{equation}
  \label{eq:fg2}
  f=-k_{\infty}^{1/2}\left(\imath_{l}A_{E} -\overline{P}_{E\, l}\right)\,.
\end{equation}

Summarizing, the condition of invariance of the gauge field $A_{E}$ under the
isometry generated by $l$ is 

\begin{equation}
-\left[  \imath_{l}F_{E}+dP_{E\, l}\right]=
-\left(\imath_{l}dA_{E}+d\imath_{l}A_{E}\right)
+d(\imath_{l}A_{E}-P_{E\, l})
=
-\pounds_{l}A_{E}+\delta_{\chi_{l}}
\equiv
-\mathbb{L}_{l}A_{E}
=
0\,,
\end{equation}

\noindent
where the ``compensating gauge transformation'' parameter $\chi_{l}$ is

\begin{equation}
  \chi_{l}
  \equiv
  \imath_{l}A_{E}-P_{E\, l}\,.
\end{equation}

$\mathbb{L}_{l}A_{E}$ is the gauge-covariant Lie (or Lie-Maxwell) derivative
of $A_{E}$ with respect to $l$ mentioned in the introduction
\cite{Ortin:2015hya,Elgood:2020svt}.\footnote{See also
  Ref.~\cite{Heusler:1993cj}.} The emergence of this formula in the KK
framework is one of our main results.

The last set of components of the 5-dimensional Killing vector equation for
$\hat{l}$ are automatically satisfied

\begin{equation}
  \pounds_{\hat{l}}\hat{g}_{\mu\nu}
  =
  \pounds_{l}g_{\mu\nu}
  -2kA_{\mu}A_{\nu}\pounds_{l}k
  -2k^{2}\left(\pounds_{l}A_{(\mu}+\partial_{(\mu}f\right)A_{\nu)}
  =
  0\,,
\end{equation}

\noindent
upon use of all the previous results.

Thus, we have constructed a 5-dimensional extension of $l$ (the
\textit{uplift} of $l$), namely

\begin{equation}
  \label{eq:hatldef}
  \hat{l}
  =
  l-k_{\infty}^{1/2}\left(\imath_{l}A_{E} -\overline{P}_{E\, l}\right)\hat{k}\,,  
\end{equation}

\noindent
which is a Killing vector of the 5-dimensional metric and whose Killing
horizon a S$^{1}$ fibration over the Killing horizon of $l$. 

On the Killing horizon itself we can write

\begin{equation}
  \hat{l}
  \stackrel{\mathcal{H}}{=}
     l -k_{\infty}^{1/2}\Omega \hat{k}\,,  
\end{equation}

\noindent
where the constant $\Omega$ is given by

\begin{equation}
  \Omega
  =
  \left.\imath_{l}A_{E}\right|_{\mathcal{H}}\,.
\end{equation}

In the solutions we are considering, $\Omega$ can be identified with the
electrostatic potential evaluated over the horizon

\begin{equation}
 \Omega = \Phi_{H}\,,  
\end{equation}

\noindent
which is a gauge-dependent quantity. Since the gauge transformations of the
4-dimensional KK vector field are 5-dimensional diffeomorphisms which are not
5-dimensional isometries, this result is not surprising. However, the
ambiguity in the value of $\Omega$ can be eliminated by demanding the
5-dimensional metric to be asymptotically flat with the following
normalization\footnote{We assume the 4-dimensional metric to be
  asymptotically-flat as well.}

\begin{equation}
  ds^{2}_{(5)}
  \longrightarrow 
  \eta_{\mu\nu}dx^{\mu}dx^{\nu} -k^{2}_{\infty}dz^{2}\,,
\end{equation}

\noindent
or, equivalently, that the KK vector field vanishes at spatial
infinity.\footnote{The vector field of the solutions
  Eqs.~(\ref{eq:4delectricsolution}) satisfy this condition.} Then,

\begin{equation}
 \Omega = \Phi\,,  
\end{equation}

\noindent
where $\Phi$ is the (gauge-invariant) difference of electrostatic potential
between the horizon and spatial infinity. Without this condition, the
coordinates $t$ and $z$ are entangled at infinity in electrically-charged
black holes, for instance.

There is another interpretation for the constant $\Omega$: the linear momentum
of free-falling observers in the direction $z$, given by

\begin{equation}
P_{z} \equiv \hat{g}_{\underline{z}\hat{\mu}}\dot{x}^{\hat{\mu}}\,,  
\end{equation}

\noindent
is a conserved quantity. When the KK vector is electric,
$\imath_{l}A= A_{t}\neq 0$, observers with $P_{z}=0$, however, are
moving in the $z$ direction with velocity

\begin{equation}
  \frac{dz}{dt}
  =
  -\imath_{l}A\,, 
\end{equation}

\noindent
which may vanish depending on the chosen gauge or, equivalently, on the chose
coordinate $z$.  This fact can be interpreted as the dragging of inertial
frames by the spacetime, which has momentum in the direction $z$. A particle
that starts at infinity with zero velocity in the compact direction and falls
radially towards the horizon will acquire a non-vanishing velocity in the
internal direction that will equal $k_{\infty}^{1/2}\Omega$ at the horizon.

This is very similar to what happens in the Kerr spacetime to zero angular
momentum observers (\textit{ZAMO}s) and, geometrically, it has to do with the
fact that the 5-dimensional vector $\partial_{t}$ is not
hypersurface-orthogonal. A difference, however, is that in these spacetimes
there may not be a static limit where $\hat{g}_{tt}\neq 0$.

A similar phenomenon happens in the magnetic case in which
$\imath_{\partial_{\varphi}}A\neq 0$. For vanishing $P_{z}$, either
$\dot{z}=\dot{\varphi}=0$ or

\begin{equation}
  \frac{dz}{d\varphi}
  =
  -\imath_{\partial_{\varphi}}A\,.
\end{equation}

To end this section, we can prove that the surface gravity of the
5-dimensional Killing horizon coincides with that of the 4-dimensional one.
First, observe that the standard definition of the 4-dimensional surface
gravity is invariant under Weyl rescalings of the metric when we write it in
the form

\begin{equation}
  \nabla_{\mu}l^{2}
  \stackrel{\mathcal{H}}{=}
  -2\kappa l_{\mu}\,,
\end{equation}

\noindent
and, therefore, we can use this definition in the Einstein or KK frames.

The 1-form dual to the Killing vector $\hat{l}$ if given by

\begin{equation}
  \hat{l}_{\hat{\mu}}dx^{\hat{\mu}}
  =
  \left[g_{t\mu} -k_{\infty}k^{2}\overline{P}_{E\, l}A_{E\, \mu}\right]dx^{\mu}
  -k_{\infty}^{1/2}k^{2}\overline{P}_{E\, l}dz\,.
\end{equation}

\noindent
It follows that 

\begin{equation}
  \hat{l}^{2}
  =
  g_{tt}-k_{\infty}k^{2}\overline{P}_{E\, l}^{2}\,,
\end{equation}

\noindent
and 

\begin{equation}
  \hat{l}_{\hat{\mu}}dx^{\hat{\mu}}
  \stackrel{\mathcal{H}}{=}
  \left. g_{t\mu}dx^{\mu}  \right|_{\mathcal{H}}
  =
  \left.l_{\mu}dx^{\mu} \right|_{\mathcal{H}}\,,
\end{equation}

\noindent
so the pullbacks of the 1-forms $\hat{l}_{\hat{\mu}}dx^{\hat{\mu}}$ and
$l_{\mu}dx^{\mu}$ are identical over the horizon even if the dual vectors are
not.  Then, on $\mathcal{H}$ only, using the vanishing of
$\overline{P}_{E\, l}$ and $g_{tt}$ there, we find

\begin{equation}
  \begin{aligned}
    \hat{\nabla}_{\mu}\hat{l}^{2}
    & =
    \nabla_{\mu}g_{tt}
    -k_{\infty}\overline{P}_{E\, l}^{2}\nabla_{\mu}k^{2}
    -2k_{\infty}k^{2}\overline{P}_{E\, l}\nabla_{\mu}\overline{P}_{E\, l}
    =
    \nabla_{\mu}g_{tt}
    =
    \nabla_{\mu}l^{2}
    =
    -2\kappa l_{\mu}
    =
    -2\kappa \hat{l}_{\mu} \,,
    \\
    & \\
    \hat{\nabla}_{\underline{z}}\hat{l}^{2}
    & =
    0
    =
    -2\kappa \hat{l}_{\underline{z}}\,,
  \end{aligned}
\end{equation}

\noindent
thus showing that the 4- and 5-dimensional surface gravities are the same.

\subsection{The electric KK black hole in 5d}
\label{sec-electric}

Using Eq.~(\ref{eq:5dimensionalmetric}) for the 4-dimensional electric
solution Eqs.~(\ref{eq:4delectricsolution}) we get the 5-dimensional
Ricci-flat and asymptotically-flat  metric

\begin{equation}
  ds^{2}_{(5)}
  =
  H^{-1}Wdt^{2}
  -W^{-1}dr^{2} - r^{2}d\Omega^{2}_{(2)}
    -H\left[d(Rz/\ell) +\alpha (H^{-1}-1)dt\right]^{2}\,.
\end{equation}

Defining the coordinates

\begin{equation}
  u
  =
  Rz/\ell-\alpha t\,,
  \hspace{1cm}
  v
  =
  \alpha t\,,
\end{equation}

\noindent
and, using the relation between the integration constants $h,\omega,\alpha$
Eq.~(\ref{eq:relation}), the metric can we rewritten in the form

\begin{equation}
  \label{eq:5delectrickkbh}
  ds^{2}_{(5)}
  =
  \frac{w}{h}\alpha^{-2}dv^{2}
  -2 du\left(dv +\tfrac{1}{2}Hdu\right)
  -W^{-1}dr^{2} -r^{2}d\Omega_{(2)}^{2}\,.
\end{equation}

\noindent
In this form we can clearly see that, in the extremal case $w=0$ (so
$W=\alpha=1$), the spacetime is a $pp$-wave with a flat 3-dimensional
wavefront

\begin{equation}
  ds^{2}_{(5)}
\,  =
  -2 du\left(dv +\tfrac{1}{2}Hdu\right)
  -\left(dr^{2} +r^{2}d\Omega_{(2)}^{2}\right)\,.
\end{equation}

These coordinates are not so useful in the non-extremal case, which is the one
we are most interested in.

The 5-dimensional metric Eq.~(\ref{eq:5delectrickkbh}) is singular at
$r=-w=r_{0}$, but this is just a coordinate singularity as can be seen by
using 4-dimensional Eddington-Finkelstein coordinates

\begin{equation}
  dt
  \equiv
  dx^{\pm} \pm H^{1/2}W^{-1} dr\,,   
\end{equation}

\noindent
together with

\begin{equation}
  d(Rz/\ell)
  \equiv
  d(R z^{\pm}/\ell) \mp  \alpha  (H^{-1}-1) H^{1/2}W^{-1} dr\,,
\end{equation}

\noindent
in which the metric takes the form 

\begin{equation}
  \begin{aligned}
    ds^{2}_{(5)}
    & =
    H^{-1}Wdx^{\pm\, 2} \pm 2 H^{-1/2}dx^{\pm}dr -r^{2}d\Omega^{2}_{(2)}
    \\
    & \\
    & \hspace{.5cm}
    -H\left[d(Rz^{\pm}/\ell)
      +\alpha (H^{-1}-1) dx^{\pm}\right]^{2}\,,
  \end{aligned}
\end{equation}

\noindent
in which all the components are regular and the determinant is different from
zero at $r=r_{0}$. Actually, according to the general arguments we have given,
there is an event horizon at $r=r_{0}$.

The 5-dimensional Killing vector field that becomes null on the event horizon
$r=r_{0}$ is just

\begin{equation}
  \hat{l}
  =
  \partial_{t} -k_{\infty}^{1/2}\Phi \partial_{\underline{z}}\,,
\end{equation}

\noindent
where $\Phi$ is the electrostatic potential evaluated on the horizon
(normalized to vanish at infinity), given in Eq.~(\ref{eq:PhiKK}).

\section{5-dimensional thermodynamics  \textit{\`a la Wald}}
\label{sec-5dthermodynamics}

In this section we want to see to which extent we can recover the
4-dimensional first law Eq.~(\ref{eq:firstlaw}) and the 4-dimensional Smarr
formula Eq.~(\ref{eq:Smarr}) working with the purely gravitational
5-dimensional action and solution  using Wald's formalism. First, we have
to find the 5-dimensional Noether-Wald charge for the 5-dimensional Killing
vector $\hat{l}$, defined in Eq.~(\ref{eq:hatldef}), whose Killing horizon
coincides with the black hole's event horizon.

The 5-dimensional Noether-Wald charge \cite{Wald:1993nt} is nothing but the
standard Komar charge 3-form \cite{Komar:1958wp}, that, for a generic Killing
vector $\hat{\xi}$, can be written in the form

\begin{equation}
  \label{eq:5dKomarcharge}
  \hat{\mathbf{Q}}[\hat{\xi}]
  =
  -\frac{1}{16\pi G_{N}^{(5)}}
  \hat{\star}(\hat{e}^{\hat{a}} \wedge \hat{e}^{\hat{b}})
  \hat{P}_{\hat{\xi}\, \hat{a}\hat{b}}\,,
\end{equation}

\noindent
where 

\begin{equation}
  \hat{P}_{\hat{\xi}\, \hat{a}\hat{b}}
  =
  \mathcal{D}_{\hat{a}}\hat{\xi}_{\hat{b}}\,,
\end{equation}

\noindent
is the \textit{Lorentz momentum map} associated to $\hat{\xi}$
\cite{Elgood:2020svt}, also known as the \textit{Killing bivector}.

Let us compare this charge with the 4-dimensional one.  The different
components of the momentum map of $\hat{l}$ can be written as follows:

\begin{subequations}
  \label{eq:momentummapreduction}
  \begin{align}
    \hat{P}_{\hat{l}\, ab}
    & =
      P_{E\, l\, ab}
      +l_{E\, [a}e_{E\, |b]}\log{k}
      -\tfrac{1}{2} k^{2}\overline{P}_{E\, l} F_{E\, ab}\,,
    \\
    & \nonumber \\
    \hat{P}_{\hat{l}\, za}
    & =
      -\tfrac{1}{2}k\, \imath_{l}F_{E\, a}
      +\overline{P}_{E\, l} e_{E\, a} k\,,
    \\
    & \nonumber \\
    \hat{P}_{\hat{l}\, az}
    & =
      -\tfrac{1}{2}k\, \imath_{l}F_{E\, a}
      -\overline{P}_{E\, l} e_{E\, a} k
      -ke_{E\, a}\overline{P}_{E\, l}\,.
  \end{align}
\end{subequations}

\noindent
Since $\hat{l}$ is, by hypothesis, a 5-dimensional Killing vector, we should
get $\hat{P}_{\hat{l}\, za}=-\hat{P}_{\hat{l}\, az}$. Indeed, this property
follows immediately from the momentum map equation.

The only term that will contribute to the integral when pulled back over
3-dimensional spacelike hypersurfaces is 

\begin{equation}
    \hat{\star}(\hat{e}^{a}\wedge \hat{e}^{b})
     =
      -\star_{E}(e_{E}^{a}\wedge e_{E}^{b})
      \wedge d(k_{\infty}z)\,.
\end{equation}

Using now this equation and the first of Eqs.~(\ref{eq:momentummapreduction})
in the Komar charge Eq.~(\ref{eq:5dKomarcharge}), we find that 

\begin{equation}
  \hat{\mathbf{Q}}[\hat{l}]
  =
  \frac{1}{16\pi G_{N}^{(5)}}
    \left[\star_{E}(e_{E}^{a}\wedge e_{E}^{b})P_{l\, ab}
      -e^{\sqrt{3}\phi}\star F_{E}\overline{P}_{E\,l}
    +\tfrac{1}{\sqrt{3}}\imath_{l}\star_{E}d\phi\right]
    \wedge d(k_{\infty}z)\,,
\end{equation}

\noindent
plus components that will vanish when pulled back over the spacelike
3-surfaces we consider. Thus, 

\begin{equation}
  \label{eq:5dversus4dNWcharges}
  \hat{\mathbf{Q}}[\hat{l}]
  =
  \frac{1}{2\pi \ell }\mathbf{Q}[l] \wedge d z\,,
\end{equation}

\noindent
where $\mathbf{Q}[l]$ is the Noether-Wald charge computed directly form the
4-dimensional action Eq.~(\ref{eq:4dNWcharge}). Integrating over $z$

\begin{equation}
\int_{0}^{2\pi \ell}   \hat{\mathbf{Q}}[\hat{l}]
  =
  \mathbf{Q}[l]\,.
\end{equation}

Let us now compare the 5- and 4-dimensional Komar charges. They are given by

\begin{subequations}
  \begin{align}
  \hat{\mathbf{K}}[\hat{l}]
  & =
    \hat{\mathbf{Q}}[\hat{l}]\,,
    \\
    & \nonumber \\
  \mathbf{K}[l]
  & =
    \mathbf{Q}[l]-\omega_{l}\,,    
  \end{align}
\end{subequations}

\noindent
where $\omega_{l}$ is given in Eq.~(\ref{eq:omegal}).  Taking into account
Eq.~(\ref{eq:5dversus4dNWcharges}) it is obvious that we are missing the terms
that give rise to the 4-dimensional 2-form $\omega_{l}$.

The apparent reason why there is a 4-dimensional $\omega_{l}$ and not a
5-dimensional $\hat{\omega}_{\hat{l}}$ is that which $\mathbf{L}\not\doteq 0$
while $\hat{\,\mathbf{L}}\doteq 0$ (see Eq.~(\ref{eq:ilL})). This is a bit
strange, because we have derived the 4-dimensional Lagrangian and equations of
motion directly from the 5-dimensional one. On closer inspection, we see that,

\begin{equation}
\mathbf{L}\doteq \tfrac{1}{\sqrt{3}}\mathbf{E}_{\phi} \doteq0\,,
\end{equation}

\noindent
and 

\begin{equation}
\imath_{l}\mathbf{L}\doteq \tfrac{1}{\sqrt{3}}\imath_{l}\mathbf{E}_{\phi}\,.   
\end{equation}

\noindent
Since, by hypothesis, $\pounds_{l}\phi=0$, we have that
$\pounds_{l}\star d \phi=0$ and using the definitions of the electric and
magnetic momentum maps we find, as in
Refs.~\cite{Pacilio:2018gom,Ballesteros:2023iqb}), that, on-shell,
$\imath_{l}\mathbf{E}_{\phi}$ is the total derivative of the closed 2-form
that whose integrals give the black-hole scalar charge $\Sigma$,
$\mathbf{Q}_{l}$. We have

\begin{equation}
  \tfrac{1}{\sqrt{3}}\imath_{l}\mathbf{E}_{\phi}
  =
\tfrac{1}{\sqrt{3}}d\mathbf{Q}_{l} 
  =
  d\omega_{l} 
 \doteq
  0\,.
\end{equation}

Thus, $\omega_{l}$ can be understood as originating in the freedom that we
have to add on-shell closed 2-forms to the Noether-Wald charge
$\mathbf{Q}[\xi]$, since we only compute directly $d\mathbf{Q}[\xi]$.

We can use this freedom directly in $\mathbf{Q}[l]$ as follows: let us
consider the exterior derivative of the last term in Eq.~(\ref{eq:4dNWcharge})
$\tfrac{1}{\sqrt{3}}\imath_{\xi}\star_{E}d\phi$ for $\xi=l$ and let us use
again the fact that, by assumption, $\pounds_{l}\star_{E}d\phi=0$. We have,
then,

\begin{equation}
  \begin{aligned}
    d\left[\tfrac{1}{\sqrt{3}}\imath_{l}\star_{E}d\phi \right]
    & =
    -\tfrac{1}{\sqrt{3}}\imath_{l}d\star_{E}d\phi =
    \tfrac{1}{\sqrt{3}}\imath_{l}\left[\mathbf{E}_{E\,\phi}
      -\tfrac{\sqrt{3}}{2}e^{\sqrt{3}\phi}F_{E}\wedge \star_{E}F_{E} \right]
    \\
    & \\
    & \doteq
    d\left\{\tfrac{1}{2}\left[e^{\sqrt{3}\phi}\star_{E}F_{E} P_{E\,l}
        +F_{E}\tilde{P}_{E\,l}\right]\right\}\,,
      \end{aligned}
\end{equation}

\noindent
after using the definitions of the momentum maps, integrating by parts and
using the Maxwell equation of motion. 

Thus, we find that

\begin{equation}
  \tfrac{1}{\sqrt{3}}\imath_{l}\star_{E}d\phi
  \doteq
    \tfrac{1}{2}\left[e^{\sqrt{3}\phi}\star_{E}F_{E} P_{E\,l}
      +F_{E}\tilde{P}_{E\,l}\right]
    +h\,,
    \,\,\,\,\,
    dh=0\,,
\end{equation}

\noindent
and, setting $h=0$ and substituting this term back into $\mathbf{Q}[l]$ in
Eq.~(\ref{eq:4dNWcharge}), we recover the Komar charge
Eq.~(\ref{eq:4dKomarcharge}).

This trick can be played in the 5-dimensional Noether-Wald charge since we
have seen that, after integration over the compact direction, it is equal to
the 4-dimensional one. In this way, we can recover the 4-dimensional Komar
charge from the 5-dimensional one and we are bound to obtain exactly the same
Smarr formula. Nevertheless, it is interesting to carry out the analysis
directly in 5 dimensions.

\subsection{5-dimensional derivation of the 4-dimensional Smarr formula}
\label{sec-Smarr}

The Smarr formula \cite{Smarr:1972kt} for 5-dimensional static,
asymptotically-flat black holes, can be obtained by integration of the
exterior derivative of the Komar charge associated to the Killing vector that
becomes null over the event horizon, $\hat{\,\mathbf{K}}[\hat{l}]$, over a
spacelike hypersurface $\Sigma^{4}$ extending from the bifurcation surface
$\mathcal{BH}$ to spacelike infinity $S^{3}_{\infty}$. Since
$\hat{\,\mathbf{K}}[\hat{l}]$ is on-shell closed

\begin{equation}
  \int_{\Sigma^{4}}d\hat{\,\mathbf{K}}[\hat{l}]
  \doteq 0\,.
\end{equation}

\noindent
On the other hand, since, by construction $\partial \Sigma^{(4)} =
\mathcal{BH} \bigcup S^{3}_{\infty}$, Stokes theorem gives

\begin{equation}
  \int_{\mathcal{BH}}\hat{\,\mathbf{K}}[\hat{l}]
  =
  \int_{S^{3}_{\infty}}\hat{\,\mathbf{K}}[\hat{l}]\,.
\end{equation}

Since, in $\mathcal{BH}$,

\begin{equation}
  \hat{P}_{\hat{l}}{}^{\hat{a}\hat{b}}
  =
  \nabla^{\hat{a}}\hat{l}^{\hat{b}}
  =
  \hat{\kappa} \hat{n}^{\hat{a}\hat{b}}\,,
\end{equation}

\noindent
where $\kappa$ is the surface gravity of the 5- and 4-dimensional Killing
horizons and $\hat{n}^{\hat{a}\hat{b}}$ is the binormal with the normalization
$\hat{n}^{\hat{a}\hat{b}}\hat{n}_{\hat{a}\hat{b}}=-2$, we have

\begin{equation}
  \int_{\mathcal{BH}}\hat{\,\mathbf{K}}[\hat{l}]
  =
  \frac{\kappa\hat{A}_{\mathcal{H}}}{8\pi G^{(5)}_{N}}\,,
\end{equation}

\noindent
where $\hat{A}_{\mathcal{H}}$ is the area of the 5-dimensional horizon and it
is related to the area of the 4-dimensional one by

\begin{equation}
  \hat{A}_{\mathcal{H} }
  =
  2\pi R A_{\mathcal{H}}\,.
\end{equation}

Then, using the relation between 5- and 4-dimensional Newton constants
Eq.~(\ref{eq:4-5Newtonconstant})

\begin{equation}
  \label{eq:integralatBH}
  \int_{\mathcal{BH}}\hat{\,\mathbf{K}}[\hat{l}]
  =
  \frac{\kappa A_{\mathcal{H}}}{8\pi G^{(4)}_{N}}
  = TS\,.
\end{equation}

\noindent
At infinity, using the boundary condition $A_{E}(\infty)=0$, we have

\begin{equation}
  \label{eq:integralatinfinity}
  \int_{S^{3}_{\infty}}\hat{\,\mathbf{K}}[\hat{l}]
  =
  \int_{S^{3}_{\infty}}\hat{\,\mathbf{K}}[\partial_{t}]
  +k_{\infty}^{1/2}\overline{P}_{E\,l\,\infty}
  \int_{S^{3}_{\infty}}\hat{\,\mathbf{K}}[\partial_{\underline{z}}]\,.
\end{equation}

\noindent
The first integral gives $2\hat{M}/3$, where $\hat{M}$ is the 5-dimensional
mass, while the second gives $2 \hat{P}_{z}/3$ where $\hat{P}_{z}$ is the
momentum of the solution along the compact dimension. Furthermore, as we have
discussed, $\overline{P}_{E\,l\,\infty}=-\Phi$.

The relation between the 5- and 4-dimensional charges can be found through the
technique of \textit{smearing}. The relation is

\begin{equation}
  \hat{M}= \tfrac{3}{4}M\,,
  \hspace{1cm}
  \hat{P}_{z}= \tfrac{3}{4}P_{z}\,.
\end{equation}

Finally, using the relation between $P_{z}$ and the 4-dimensional charge $q$
associated to the KK vector field $A_{E}$, Eq.~(\ref{eq:chargeversusmomentum})
we can rewrite the integral at infinity  Eq.~(\ref{eq:integralatinfinity}) as

\begin{equation}
  \label{eq:integralatinfinity2}
  \int_{S^{3}_{\infty}}\hat{\,\mathbf{K}}[\hat{l}]
  =
(M-\Phi q)/2\,,
\end{equation}

\noindent
and equating this result to that of the integral over the bifurcation surface,
Eq.~(\ref{eq:integralatBH}) we recover the 4-dimensional Smarr formula
Eq.~(\ref{eq:Smarr}). Observe that, in this framework, it does not seem
possible to obtain the magnetic terms unless we introduce in a more or less
arbitrary form the total derivative we discussed above.

We can also try to derive the 4-dimensional version corresponding to the KK
theory of the first law of black hole mechanics using the 5-dimensional
Noether-Wald charge, since we have shown that it can be dimensionally reduced
to the 4-dimensional one. However, as we are going to show, this is an
entirely different calculation in which there are many subtleties.

\subsection{5-dimensional derivation of the 4-dimensional first law}
\label{sec-firstlaw}

The 5-dimensional first law can be derived from the identity
\cite{Wald:1993nt,Ortin:2022uxa,Ballesteros:2023iqb}:

\begin{equation}
  d\hat{\mathbf{W}}[\hat{l}]
  \doteq
  0\,,
\end{equation}

\noindent
where

\begin{equation}
\hat{\mathbf{W}}[\hat{l}]
  \equiv
  \delta\hat{\mathbf{Q}}[\hat{l}]
  +\imath_{\hat{l}}\hat{\mathbf{\Theta}}(\hat{e},\delta\hat{e})
  -\varpi_{\hat{l}}\,,
\end{equation}

\noindent
where $\hat{\mathbf{\Theta}}(\hat{e},\delta\hat{e})$ is the
\textit{presymplectic 3-form} defined in Ref.~\cite{Lee:1990nz} and
$\varpi_{k}$is implicitly defined by\footnote{The origin of this term is the
  induced local Lorentz transformations \cite{Ortin:2022uxa}.}

\begin{equation}
  \label{eq:varpidef}
  \delta_{\hat{\sigma}_{\hat{l}}}\hat{\mathbf{\Theta}}(\hat{e},\delta\hat{e})
  \equiv
  d\varpi_{\hat{l}}\,.
\end{equation}

\noindent
Furthermore, it is assumed that the variations of the Vielbeins $\delta e$
satisfy the linearized equations of motion in the black-hole's background.

In Ref.~\cite{Ortin:2022uxa} it was found that, once all these details have
been taken into account, $\hat{\mathbf{W}}[\hat{l}]$ takes the
form\footnote{It ihas to be taken into account that
  $\mathbf{W}=-\mathbf{\Omega}$ in Ref.~\cite{Ortin:2022uxa}.}

\begin{equation}
  \hat{\mathbf{W}}[\hat{l}]
  =
  -\hat{P}_{\hat{l}\, \hat{a}\hat{b}}\delta \hat{\star}
  (\hat{e}^{\hat{a}}\wedge \hat{e}^{\hat{b}})
  -\imath_{\hat{l}}\hat{\star} (\hat{e}^{\hat{a}}\wedge \hat{e}^{\hat{b}})
  \wedge\delta \hat{\omega}_{\hat{a}\hat{b}}\,.
\end{equation}

This identity is to be integrated over the same spacelike hypersurface we used
to integrate the derivative of the Komar charge, leading to 

\begin{equation}
  \int_{S^{3}{}_{\infty}}\hat{\mathbf{W}}[\hat{l}]
  =
  \int_{\mathcal{BH}}\hat{\mathbf{W}}[\hat{l}]\,.
\end{equation}

In this case, since we are dealing with the Einstein-Hilbert action, following
Ref.~\cite{Iyer:1994ys}, almost by definition, the first integral simply gives
the variation of the conserved charges associated to the Killing vector
$\hat{l}$, that is, a linear combination of the variation of the mass and the
variation of the momentum in the $z$ direction, which is, essentially, the
4-dimensional electric charge. The second integral, on the other hand, gives
the surface gravity times the variation of the 5-dimensional area, which is
related to the 4-dimensional one as we have shown above.

\begin{subequations}
  \begin{align}
    \int_{S^{3}{}_{\infty}}\hat{\mathbf{W}}[\hat{l}]
    & =
      \delta \hat{M} -k_{\infty}^{1/2}\Phi \delta \hat{P}_{z}\,,
    \\
    & \nonumber \\
    \int_{\mathcal{BH}}\hat{\mathbf{W}}[\hat{l}]
    & =
    \frac{\kappa}{8\pi G^{(5)}_{N}} \delta \hat{A}_{\mathcal{H}}\,.
  \end{align}
\end{subequations}

In principle, there is no work term associated to the variation of the modulus
$\phi_{\infty}$ but, since some of the 5- and 4-dimensional variables (charge,
electrostatic potential, horizon area) are related by factors that involve the
modulus $\phi_{\infty}$ and, therefore, their variations lead to an extra term
involving the variation of the modulus.\footnote{Remember that the
  compactification radius $R$ is invariant but $\ell$ and, hence,
  $k_{\infty}$, are not.}

All this is qualitatively correct, but we have not been able to recover the
4-dimensional first law with the correct numerical coefficients. We believe
that a more detailed study of the definition of mass with KK asymptotics, is
necessary and work in this direction is already underway \cite{kn:G-FOZ}.


\section{Discussion}
\label{sec-discussion}

In this paper we have investigated the 5-dimensional geometry, spacetime
symmetries and thermodynamics of 4-dimensional, static, KK black holes. In
particular, we have determined the existence of a 5-dimensional stationary
event horizon associated to the 4-dimensional static one (more precisely, a
U$(1)$ fibration over it) and we have shown how the 4-dimensional Killing
vectors can be uplifted to 5-dimensional Killing vectors including a piece
proportional to the generator of translations in the compact dimension. In
other words: as stated in the introduction 4-dimensional spacetime isometries
induce gauge transformations. The natural emergence of the covariant Lie
derivative (here Lie-Maxwell derivative) in this context is quite remarkable.

Applying the relation between 4- and 5-dimensional Killing vectors to the
generator of 4-dimensional time translations we have obtained a Killing vector
which is null on the 5-dimensional event horizon, proving that it is a Killing
horizon as well. This allows us to use all the geometrical properties enjoyed
by Killing horizons. In particular, we have used the standard mathematical
definition of surface gravity to show that those of the 4- and 5-dimensional
horizons must be equal. On the other hand, the interpretation of the
4-dimensional electrostatic potential on the horizon as the velocity in the
internal direction with which a particle with zero momentum in that direction
(hence, zero electric charge in 4 dimensions) crosses the horizon provides a
very interesting alternative physical interpretation of this quantity.

Our success in relating the 4- and 5-dimensional thermodynamics has not been
complete, however. A more thorough study of how the definitions of
gravitational conserved charges in \cite{Iyer:1994ys} applies to spacetimes
with KK asymptotics is necessary. Furthermore, the answer to the question of
how terms proportional to the variation of the 4-dimensional magnetic charge
can be obtained in the first law directly from 5-dimensional expressions is
unclear. The 5-dimensional backgrounds that give rise to 4-dimensional ones
with magnetic charges do not asymptote to Ricci-flat metrics
\cite{Bombelli:1986sb,Deser:1988fc} and the formalism has to be revised in
detail \cite{kn:G-FOZ}.

It is also unclear how 4-dimensional scalar charges may arise in the
5-dimensional framework. Recently, we have found a coordinate-independent
definition for these (non-conserved!) charges that satisfies a Gauss law
\cite{Ballesteros:2023iqb} and which is related to the global scaling symmetry
of the 4-dimensional theory. The 5-dimensional generator of this global
symmetry is the vector $z\partial_{\underline{z}}$ which does not leave
invariant the 5-dimensional metrics under consideration and, therefore, there
is no 5-dimensional gravitational conserved charge associated to it, but we do
not know how to construct another charge that satisfies a Gauss law even if it
is not conserved.  Given the ubiquity of scalar fields in KK theories, a
deeper understanding of this problem and of the meaning of conserved charges
is desirable. Work in this direction is in progress Ref.~\cite{kn:BG-FOZ}.

\section*{Acknowledgments}

The authors would like to thank David Pere\~n\'{\i}guez for drawing our
attention to this problem and for many useful conversations.  This work has
been supported in part by the MCI, AEI, FEDER (UE) grants PID2021-125700NB-C21
(``Gravity, Supergravity and Superstrings'' (GRASS)), PID2021-123021NB-I00 and
IFT Centro de Excelencia Severo Ochoa CEX2020-001007-S and by FICYT through
the Asturian grant SV-PA-21-AYUD/2021/52177 The work of CG-F was supported by
the MU grant FPU21/02222. The work of MZ was supported by the fellowship
LCF/BQ/DI20/11780035 from ``la Caixa'' Foundation (ID 100010434). TO wishes to
thank M.M.~Fern\'andez for her permanent support.

\appendix


\end{document}